\documentclass[a4paper,fleqn,usenatbib]{mnras}
\pdfoutput=1

\usepackage[T1]{fontenc}
\usepackage{ae,aecompl}

\usepackage{graphicx}	
\usepackage{amsmath}	
\usepackage{amssymb}	
\usepackage{newtxtext,newtxmath}

\title[Seismic modelling of KIC\,8264293]{Seismic modeling of a very young SPB star -- KIC\,8264293}

\author[Szewczuk, Walczak, Daszy{\'n}ska-Daszkiewicz \& Mo{\'z}dzierski]{Wojciech Szewczuk\thanks{E-mail: wojciech.szewczuk@uwr.edu.pl (WS)},
Przemys{\l}aw Walczak,
Jadwiga Daszy{\'n}ska-Daszkiewicz,
\newauthor
Dawid Mo{\'z}dzierski
\\
University of Wroc{\l}aw, Faculty of Physics and Astronomy, Astronomical Institute, Kopernika 11, PL-51-622 Wroc{\l}aw, Poland
}

\date{Accepted XXX. Received YYY; in original form ZZZ}

\pubyear{2015}

\begin{document}
\label{firstpage}
\pagerange{\pageref{firstpage}--\pageref{lastpage}}
\maketitle

\begin{abstract}
KIC\,8264293 is a fast rotating B-type pulsator observed by Kepler satellite.
Its photometric variability is mainly due to pulsations in high-order g modes. Besides, we detected
a weak H$\alpha$ emission.  Thus, the second source of variability 
is  the fluctuation in a disk around the star.
The pulsational spectrum of KIC\,8264293 reveals a frequency grouping and period spacing pattern.
Here, we present the thorough seismic analysis of the star based on these features.
Taking into account the position of the star in the HR diagram and fitting the 14 frequencies that form
the period spacing we constrain the internal structure of the star.
We conclude that the star barely left the ZAMS 
and the best seismic model has $M = 3.54\,\mathrm{M}_{\sun}$, $V_\mathrm{rot}=248\,\mathrm{km\,s}^{-1}$
and  $Z = 0.0112$. We found the upper limit on the mixing at the edge of the convective core,
with the overshooting parameter up to $f_\mathrm{ov} = 0.03$.
On the other hand, we were not able to constrain the envelope mixing for the star.
To excite the modes in the observed frequency range,  we had to modify  the opacity data. Our best seismic model with
an opacity increase
by 100\% at the "nickel" bump $\log T=5.46$ explains the whole instability.
KIC\,8264293 is the unique, very young  star pulsating in high-order g modes with the Be feature.
However, it is not obvious whether the source of this circumstellar matter is the ejection of mass from the underlying star
or whether the star has retained its protostellar disk.

\end{abstract}




\section{Introduction}
\label{sec:intro}
Pulsations are very common among stars of the B spectral type.
The two mostly studied classes of B-type pulsators are $\beta$ Cephei stars and Slowly Pulsating B-type (SPB) stars.
The oscillation spectra of  $\beta$ Cephei stars are dominated by the low-order p/g modes \citep[e.g.\,][]{2005ApJS..158..193S}
whereas 
frequencies of SPB stars are associated with high-order g modes \citep{1991A&A...246..453W}.
In fact, in the age of space observations, it is difficult to talk about these classes of stars separately
because in most B-type main sequence stars, if not all, variability is detected both in the p-mode as well as g-mode regime
\citep[e.g.\,][]{2005MNRAS.360..619J, 2006MNRAS.365..327H, 2006A&A...448..697C, 2011MNRAS.413.2403B}.

A separate group is formed by B-type stars showing emission in the Balmer and some metalic lines, i.e, Be stars
\citep{1979SSRv...23..541S, 2003PASP..115.1153P}.
The classical Be stars are main-sequence stars or subgiants, that rotate rapidly and the emission
originates in a circumstellar disk \citep[e.g.,][]{2013A&ARv..21...69R}.
Because the classical Be star are too old to have retained the protostellar disk,
the source of this circumstellar matter must be from the ejection ({\it decretion disk}) of mass from the underlying
star. Although the exact mechanism of the disk formation is still a puzzle underlying  the Be phenomenon
\citep[e.g.,][]{2013A&ARv..21...69R},
it is generally believed that the combined effects of very  rapid rotation and non-radial pulsations trigger the mass ejection
\citep[e.g.,][]{1986PASP...98...30O,1988IAUS..132..217B,2005ASPC..337..101O,2017sbcs.conf..196B,2020A&A...644A...9N}.
However, a competing hypothesis was recently proposed by \citet{2020MNRAS.493.2528B}
who suggested that magnetic activity leads to flaring outburst  and ejection of material.

There is also a group of pre-main sequence stars with masses in the range of about 2--8$M_{\sun}$ , i.e., the Herbig Be stars,
that exhibit  similar spectroscopic features to the classical Be stars
\citep[e.g.,][]{1960ApJS....4..337H, 1998ARA&A..36..233W}.

On the other hand, in a very young B-type star, the disk may have retained after the protostar phase ({\it accretion disk})
with the residual emission in the Balmer lines and infrared. Finally,  one can also presume that a star can just enter the classical Be phase
and shows a very week emission.

Detection of pulsations in any class of B-type stars allow for testing various hypotheses  of their formation
and for constraining their internal structure. In particular, high-order g modes are very sensitive to the physical conditions
near the convective core. Thus, seismic modelling based on gravity modes gives a great opportunity to obtain constraints
on internal mixing processes and angular momentum transport mechanism.

On the one hand, internal chemical mixing in stars is still a source of uncertainty in our evolutionary calculations.
On the other hand, it has a huge impact on stellar evolution \citep{2009pfer.book.....M, 2012sse..book.....K}.
In addition to extending the main-sequence lifetime,
chemical mixing also substantially increases the mass of the stellar core 
and causes the star to evolve at higher  luminosities in subsequent evolutionary phases.


There are many interesting results from seismic modelling  of  B-type stars
\citep[e.g.\,][]{2012A&A...539A..90N, 2013MNRAS.431.3396D, 2015A&A...580A..27M, 2015ApJ...810...16T,
2015MNRAS.453..277S, 2016ApJ...823..130M,
2017EPJWC.16003012S, 2018MNRAS.478.2243S, 2019A&A...632A..95M,
2020A&A...644A...9N,2021NatAs.tmp...80P}  but such studies are often inhibited by  the lack of mode identification.

In the case of high-order g modes the asymptotic theory predicts, so-called, (quasi-)regular period spacing.
These quasi-equally spaced in period structures consist of  modes with consecutive radial orders, $n$, of the same harmonic degree, $\ell$, and
azimuthal order, $m$ \citep[e.g.\,][]{1980ApJS...43..469T, 	1993MNRAS.265..588D, 2013MNRAS.429.2500B, 2017MNRAS.469...13S}.
Such characteristic patterns in oscillation spectrum allow for mode identification.
However, their observational discovery had to wait until the era of space borne high precision high cadence photometric times series
\citep[e.g.\,][]{2012A&A...542A..88D, 2014A&A...570A...8P, 2015ApJ...803L..25P, 2017A&A...598A..74P, 2018MNRAS.478.2243S, 
2021MNRAS.503.5894S, 2021NatAs.tmp...80P}.

Another problem in seismic modeling of B-type stars is the fairly common problem with the excitation of high-order g modes
\citep[e.g.,][]{2004MNRAS.350.1022P}.
Asteroseismic studies of B-type pulsators \citep[e.g.,][]{2017sbcs.conf..173W, 2017EPJWC.15206005W, 2017sbcs.conf..138D,2017MNRAS.466.2284D,
2018MNRAS.478.2243S, 2019MNRAS.485.3544W} imply that maybe the stellar opacities still requires some revision.
In particular, it was shown that a huge increase of opacity near the Z-bump is necessary to explain low-frequency  modes observed in the early  B-type stars \citep{2012MNRAS.422.3460S,2017MNRAS.466.2284D}.
On the other hand, the accumulation of heavy elements in the driving layers can significantly increases the opacity
and contribute to the mode excitation. Recently, \citet{2018A&A...610L..15H} in their theoretical work
showed, that including diffusion  does cause the accumulation of iron and nickel
in the vicinity of the driving zone of  massive B-type stars.

In this paper, we present seismic modelling of the fast rotating SPB star KIC\,8264293 which appeared to be a very young object
with a weak emission in the H$\alpha$ line.
In particular, we try to obtain constraints on the mixing efficiency near the convective core and in the radiative envelope.

In Section \ref{sec:KIC8264293}, we give  details about the stellar parameters of  KIC\,8264293 and its Be feature.
Section \ref{Sect:sm} contains  description and results of our seismic modelling with the standard opacity data.
In Section \ref{Sect:mod_opac} we explore modifications of stellar opacities and their impact on the mode excitation.
Conclusions are summarized in Section \ref{conclusions}.

\section{KIC\,8264293}
\label{sec:KIC8264293}

\subsection{Basic stellar parameter and emission in H$\alpha$}

KIC\,8264293 is a B-type star with a brightness of $V=11.264(1)$\,mag \citep{2015MNRAS.453.1095B}.
From spectroscopic observations \citet{2019AJ....157..129H} determined the effective temperature,
$T_\mathrm{eff}=13150(500)$\,K, surface gravity,  $\log g=4.08(14)$ and the projected rotation velocity, $V\sin i=284(13)\,\mathrm{km\,s^{-1}}$
which is about 60\% of the break-up velocity.
Its reddening, as determined from Bayestar2019 extinction maps \citep{2019ApJ...887...93G}, is $E(B-V)=0.1407(79)$\,mag. This translates into extinction $A_V=0.436(37)$.
Based on the Gaia EDR3 parallax $\pi = 0.655(16)$ \citep{2020arXiv201201533G} we found
$\log L/\mathrm L_{\sun}=2.295(41)$. Here, we used the bolometric correction $BC=-0.914(75)$
from \citet{1996ApJ...469..355F}.  The values in the parentheses refers to the errors
on the parameters.

\begin{figure}
	\includegraphics[angle=0, width=\columnwidth]{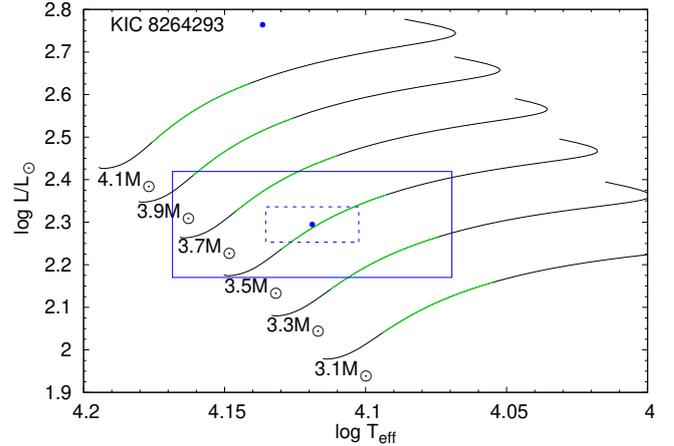}
    \caption{The H-R diagram with the position of KIC\,8264293 (blue dot). The rectangles marked with blue dashed and solid lines represent $1\sigma$ and
    $3\sigma$ error boxes, respectively. Evolutionary tracks are plotted for masses in the range from $3.1\,\mathrm{M}_{\sun}$ to $4.1\,\mathrm{M}_{\sun}$
    and were calculated with MESA code for $X_0=0.71$, $Z=0.014$, $f_\mathrm{ov}=0.02$ and the OPLIB opacity tables. Zero rotation was assumed.
    Green parts of the tracks correspond to models that reproduce the observed surface gravity within $1\sigma$ error.
    }
    \label{fig:HR_KIC8264293}
\end{figure}

The position of the star on the Hertzsprung-Russell (HR) diagram is shown in
Fig.\,\ref{fig:HR_KIC8264293}. In addition we marked evolutionary tracks for
masses from $3.1\,\mathrm M_{\sun}$ to $4.1\,\mathrm M_{\sun}$ with a step
$0.2\,\mathrm M_{\sun}$. Initial hydrogen abundance,
$X_0=0.71$, metallicity, $Z=0.014$, and overshooting parameter in exponential
description,
$f_\mathrm{ov}=0.02$ were assumed. As a source of the opacity data
we used OPLIB \citep{2015HEDP...14...33C,2016ApJ...817..116C}
tables. All effects of rotation on stellar evolution were neglected.
The evolutionary  models were calculated with the use of
Modules for Experiments in Stellar Astrophysics code version 9575
\citep[MESA][]{Paxton2011, Paxton2013, Paxton2015, Paxton2018, Paxton2019}.
The code uses equation of state, which is a blend of the OPAL \citet{Rogers2002}, SCVH
\citet{Saumon1995}, PTEH \citet{Pols1995}, HELM
\citet{Timmes2000}, and PC \citet{Potekhin2010} EOSes. The OPLIB tables
of radiative opacities were supplemented with low-temperature data from \citet{Ferguson2005}
and the high-temperature, Compton-scattering dominated regime by
\citet{Buchler1976}.  Electron conduction opacities are from
\citet{Cassisi2007}. Nuclear reaction rates are a combination of rates from
NACRE \citep{Angulo1999}, JINA REACLIB \citep{Cyburt2010}, plus additional tabulated weak reaction rates \citet{Fuller1985, Oda1994, Langanke2000}.  Screening is
included via the prescriptions of \citet{Salpeter1954, Dewitt1973,
Alastuey1978, Itoh1979} and  thermal
neutrino loss rates are from \citet{Itoh1996}.

\begin{figure}
	\includegraphics[angle=270,width=\columnwidth]{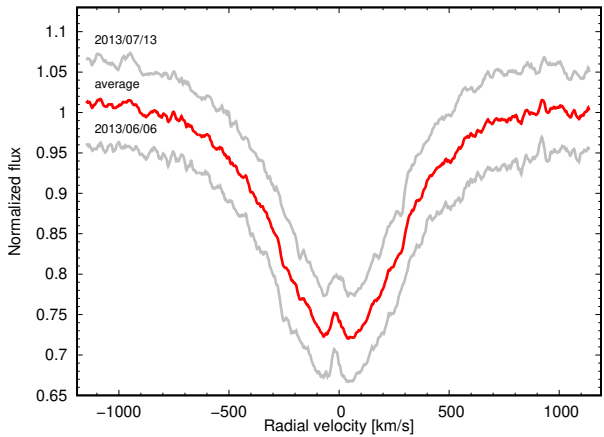}
    \caption{The two normalized line profiles of H$\alpha$ (the grey lines) of KIC\,8264293 and the average line profile (the red line).
    For a better visibility the individual spectra were shifted in the flux by 0.05.}
    \label{Ha}
\end{figure}

KIC\,8264293 lies in the field of the open cluster NGC\,6866. The membership probability
was estimated by several author and it is $79\%$  according to \citet{2008AJ....136..118F},
$90\%$ according to \citet{2009AcA....59..193M} and $43\%$ as given by \citet{2015MNRAS.453.1095B}.
On the other hand the distance to the KIC\,8264293 as determined from the GAIA EDR3
parallax, 1526(38) pc, places  the star in the background of the closer NGC\,6866.
The distance to this cluster is 1392(44)\,pc as estimated by \citet{2020MNRAS.499.1874M} and  1189(75)\,pc
according to \citet{2015MNRAS.453.1095B}. Thus, the star is most probably not a member of the cluster.

The star was  observed by one of us at the Apache Point Observatory with ARC 3.5-m telescope.
We took the two spectra with the use of ARC \'{E}chelle Spectrograph (ARCES) that has the resolving power  
of 31\,500  and covers the spectra range of 320--1000 nm.
The spectra were taken during two nights, June 6 and July 13, 2013, both with 1800 s exposures.
We have detected a weak emission in the core of the H$\alpha$ spectral line. Fig.~\ref{Ha} shows these two H$\alpha$ profiles, obtained
about one month apart, as well as the average profile with the signal-to-noise ratio of about 130. Although the emission profile is relatively small,
it seems that it has changed a little during this period. Our spectra indicate that KIC\,8264293 is a Be-type star. Such stars show emission in a wide range of intensities \citep[see a comprehensive review by][]{2013A&ARv..21...69R}, including phases with no emission at all, during disk-less episodes. Switching between emission and disk-less episodes were detected e.g. for stars in NGC\,457 open cluster by \citet{2014AcA....64...89M}.

\begin{figure*}
	\includegraphics[angle=0, width=2\columnwidth]{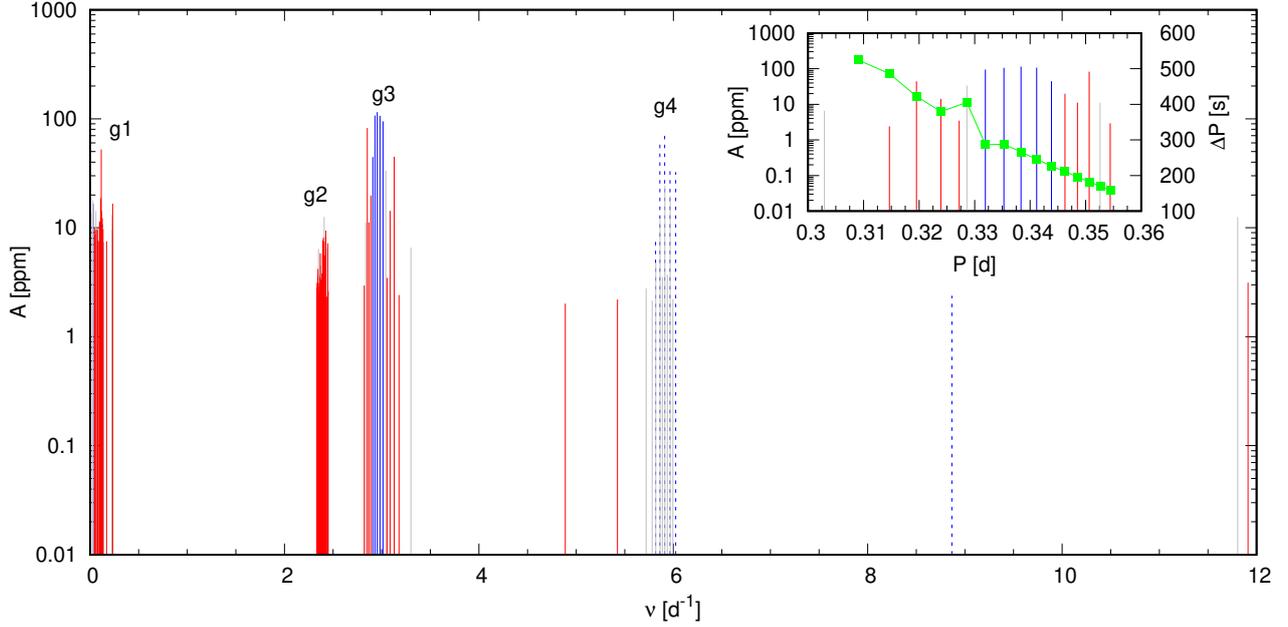}
    \caption{The oscillation spectrum of KIC\,8264293. Independent frequencies are marked with vertical red lines
    and combinations with vertical gray lines. Vertical  blue solid lines indicate independent frequencies for which harmonics are present.
    Harmonics  are marked
    with vertical blue dashed lines. The inset shows a zoom-in on the g3 group  (left y-axis) and detected period spacing (green squares and green line, right y-axis)
    as a function of period. Errorbars are smaller than symbol size.
    }
    \label{fig:KIC8264293_osc_spec}
\end{figure*}


\begin{figure*}
	\includegraphics[angle=270, width=\columnwidth]{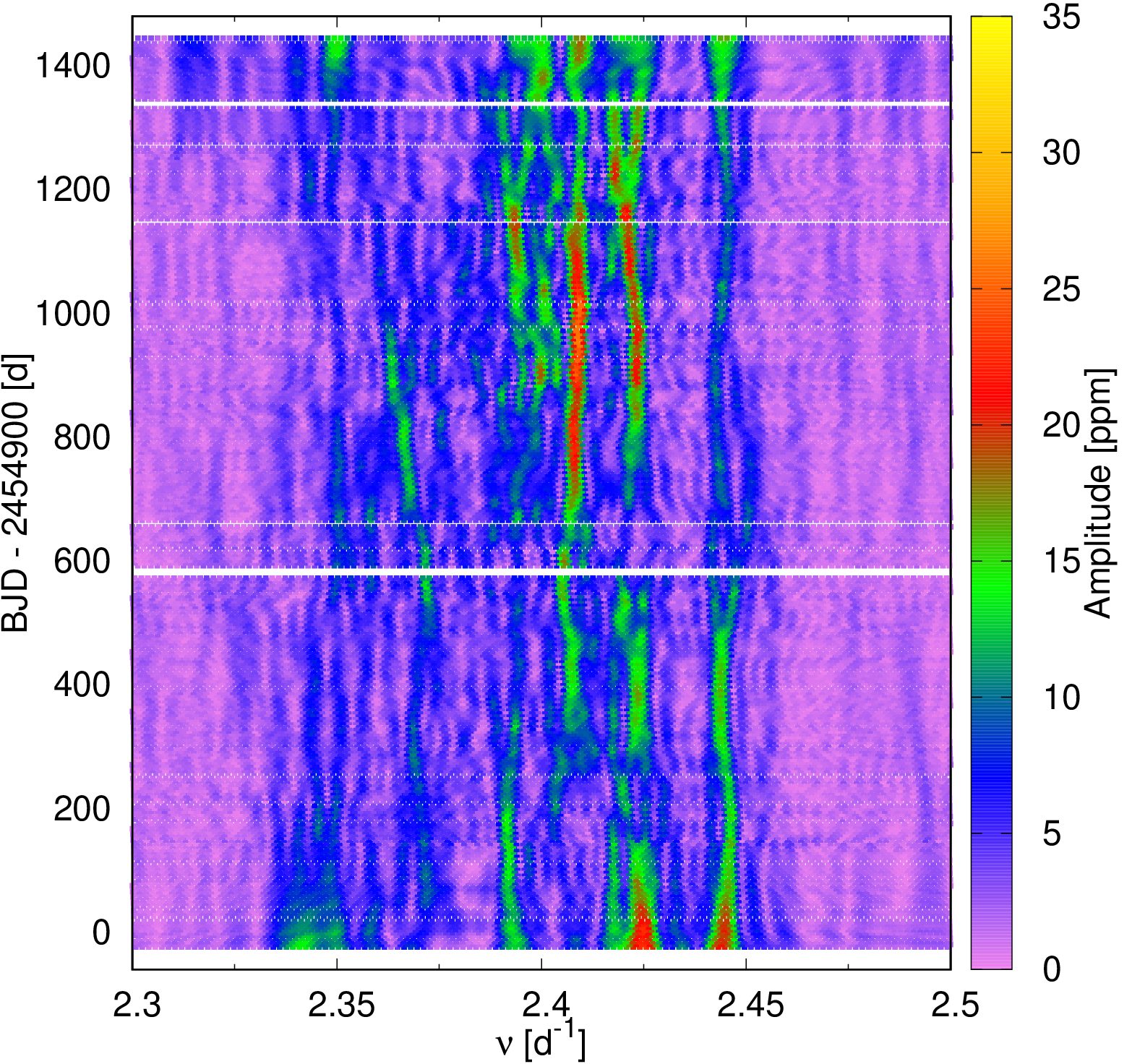}
	\includegraphics[angle=270, width=\columnwidth]{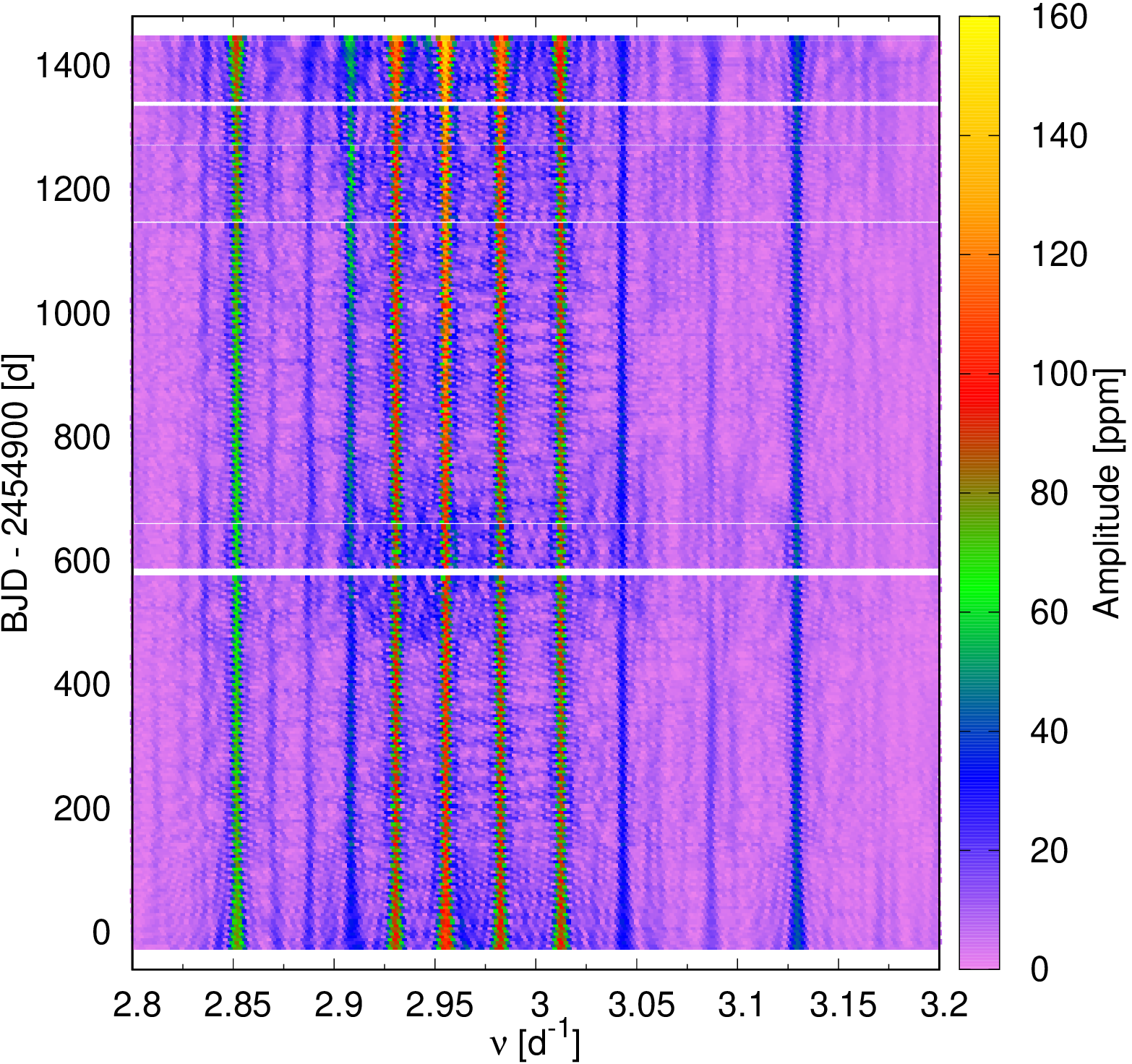}
    \caption{Time-depended Fourier amplitude spectrum calculated
    for the data after prewithening on 42 frequencies in the vicinity of
    g2 group of frequencies (left panel) and calculated
    for the original
    data in the vicinity of g3 group of frequencies (right panel).
    The amplitude values are coded with colours.}
    \label{fig:TD_osc_spec}
\end{figure*}

\setlength{\tabcolsep}{4.5pt}
\begin{table}
	\centering
	\caption{The frequencies forming the regular period spacings found by \citet{2021MNRAS.503.5894S} in the
	oscillation spectrum of KIC\,8264293. In the following columns there are given: the frequency id (ID), frequency value ($\nu$), period (P),
period difference ($\Delta P$), amplitude (A) and signal to noise ratio (S/N).
	In the last column there is given the frequency type (ft), 'i' stands for an independent frequency and 'c' for a combination frequency.}
	\label{tab:S8264293}
	\begin{tabular}{rrrrrrr} 
		\hline
 ID          &    $\nu$      & $P$    &$\Delta P$& $A$     & $\frac{\mathrm S}{\mathrm N}$         &   ft\\
             &    $(\mathrm{d}^{-1})$ & $(\mathrm{d})$  &$(\mathrm{d})$  & $(\mathrm{ppm})$    &               &  \\
\hline
$\nu_{133}$ & 2.82140(9)  & 0.35444(1)  & 0.00183(1)  & 3.0(4)   & 5.1 & i \\
$\nu_{40}$  & 2.83603(3)  & 0.352606(4) & 0.001966(4) & 11.1(4)  & 13 & c \\
$\nu_{6}$   & 2.851930(6) & 0.350639(1) & 0.002106(4) & 82(2)    & 52 & i \\
$\nu_{41}$  & 2.86917(3)  & 0.348533(4) & 0.002265(4) & 11.2(4)  & 14 & i \\
$\nu_{19}$  & 2.88794(2)  & 0.346267(2) & 0.002443(2) & 19.7(4)  & 23 & i \\
$\nu_{10}$  & 2.90846(1)  & 0.343824(1) & 0.002634(1) & 44(1)    & 43 & i \\
$\nu_{2}$   & 2.930918(6) & 0.341190(1) & 0.002844(1) & 107.4(9) & 49 & i \\
$\nu_{1}$   & 2.955556(6)  & 0.338346(1) & 0.003078(1) & 115.0(9) & 48 & i \\
$\nu_{3}$   & 2.982687(6) & 0.335268(1) & 0.003330(1) & 106.3(9) & 57 & i \\
$\nu_{4}$   & 3.012605(6) & 0.331939(1) & 0.003323(1) & 94.8(4)  & 60 & i \\
$\nu_{11}$  & 3.04307(1)  & 0.328615(1) & 0.004685(3) & 33.4(4)  & 43 & c \\
$\nu_{27}$  & 3.08709(2)  & 0.323930(3) & 0.004388(3) & 14.3(4)  & 22 & i \\
$\nu_{9}$   & 3.12948(1)  & 0.319542(1) & 0.00489(1)  & 44.7(4)  & 44 & i \\
$\nu_{156}$ & 3.1781(1)   & 0.31465(1)  &             & 2.4(4)   & 4.9 & i \\
\hline
	\end{tabular}
\end{table}

\subsection{Photometric variability}

The Kepler photometry of KIC\,8264293 has been recently analyzed by \citet{2021MNRAS.503.5894S}.
The authors used almost continues Kepler data from quarters Q0--Q17 spanning 1470 days what results in the Rayleigh resolution of 0.00068\,d$^{-1}$.
The authors found a rich oscillation spectrum that show clear frequency grouping. They found 62 independent frequencies, 33 combinations
and 7 harmonics. The whole oscillation spectrum of KIC\,8264293 is depicted in Fig.\,\ref{fig:KIC8264293_osc_spec}.
In the Kepler light curve of KIC\,8264293 we do not find signs of outbursts that are known to occur in Be stars
\citep[e.g.][]{2017A&A...598A..74P}.

As one can seen there are four distinct groups of frequencies. Moreover, in the group marked by g3 in Fig.\,\ref{fig:KIC8264293_osc_spec},
containing the highest amplitude frequencies, \citet{2021MNRAS.503.5894S} found a regular period spacing, i.e.,  quasi-equally spaced periods.
The frequencies from this sequence are given in Table\,\ref{tab:S8264293}, where we gave also the frequency ID \citep[as given in ][]{2021MNRAS.503.5894S}, the frequency value,
period, period difference, amplitude, signal-to-noise ratio and frequency type (i means an independent frequency and c - a combination).
However, in the table we have omitted the last frequency, $\nu_{86}=3.30087(4)\,\mathrm{d}^{-1}$, given by
\citet{2021MNRAS.503.5894S} because it is preceded by the missing frequency (i.e. not detected in the data).
Period spacing corresponding to the $i$-th frequency is defined as $\Delta P_i=P_{i+1}-P_i$.
From the shape of a function $\Delta P(P)$ it is possible to deduce the mode degree, $\ell$, and  azimuthal order, $m$  \citep[e.g.\,][]{2017MNRAS.469...13S}.

According to \citet{2021MNRAS.503.5894S}, the g4 group  consists of only combinations and harmonics.
However, this may be due to accidentally meeting the criteria for identifying frequencies as combinations or harmonics.
Our seismic modeling in the following sections shows that this may be the case.

The g1 group contains the lowest frequencies with the values below $0.24\,\mathrm{d}^{-1}$.
Keeping in mid their relatively low values, even $0.02472(3)\,\mathrm{d}^{-1}$ in the case of $\nu_{30}$,
they may be related to: pulsations, long term fluctuations in the disk or instrumental effects.

The nature of a variability associated with the g2 group can be deduced from the  left panel of
Fig.\,\ref{fig:TD_osc_spec} where we plotted
the time-depended Fourier amplitude spectrum in the frequency range from 2.3\,d$^{-1}$
to 2.5\,d$^{-1}$. The individual Fourier spectra that make up the 
time-depended Fourier spectrum were calculated for data subsets with a width of 300 days.
It is clear that the g2 group consists  mainly of the incoherent variability.
For a comparison, one can see a coherent behaviour of the frequencies from the g3 group in the right panel
of Fig.\,\ref{fig:TD_osc_spec}.
It is worth to add that the  rotational frequency  is just in the range (2.3,\,2.5) \,d$^{-1}$
and the frequencies of  the g2 group exhibit very strong amplitude changes (cf. the left panel of Fig.\,\ref{fig:TD_osc_spec}).

Taking into account a Be-nature of the star, incoherent signals seen in g2 group  may be identified as
so-called \u Stefl frequencies connected with a disc around the star. They are  sometimes detected during
outburst events in Be stars  and are interpreted as tracing large-scale gas-circulation \citep[e.g.][]{2000ASPC..214..240S, 2016A&A...588A..56B, 2018AJ....155...53L}.

\u Stefl frequencies are typically slightly lower than nearby pulsational frequencies. In the case of KIC\,8326493  peaks in
the g2 group have frequencies about 20\% lower than peaks in g3 group. This is the same case as in other two Be stars,
$\eta$ and $\mu$ Centauri  that have \u Stefl frequencies  about 10–20\% lower than a nearby strong pulsational frequency
\citep[][]{2016A&A...588A..56B}.

Finally, in Fig.\,\ref{fig:KIC8264293_osc_spec} one can see five separated frequencies.
Three of them, $\nu_{171}= 4.8894(1)\,\mathrm{d}^{-1}$, $\nu_{165}=5.4258(1)\,\mathrm{d}^{-1}$ and
$\nu_{129}=11.91170(9)\,\mathrm{d}^{-1}$ are identified as independent ones, while
$\nu_{155}=8.8667(1)\,\mathrm{d}^{-1}$ is a harmonic and $\nu_{33}=     11.80533(3) \,\mathrm{d}^{-1}$ is a combination.

Occurrence of the Be phenomenon and nonradial pulsations in  KIC\,8326493
may be another indication that pulsations play a role in disk formation around
Be stars. However, the study of this problem is beyond the scope of this paper.

\section{Seismic modeling}
\label{Sect:sm}

The aim of this section is to find seismic models that reproduce the regular period spacing found
in the oscillation spectrum of KIC\,8264293 (i.e. frequencies listed in Table\,\ref{tab:S8264293}).
Without an independent information on $\ell$ and $m$, this has to be done simultaneously with the mode  identification.
According to the asymptotic theory of oscillation, modes with the given values of ($\ell,~m$ ) and consecutive radial orders, $n$,
form (quasi-)regular period spacing \citep[e.g.\,][]{1980ApJS...43..469T, 1993MNRAS.265..588D, 2013MNRAS.429.2500B}. Therefore,
under the assumption that frequencies that form period spacing pattern in the g3 group
(see Table\,\ref{tab:S8264293}) have the same
mode degrees, the same azimuthal orders and consecutive radial orders, we looked for models
that  reproduce them best.
In this paper the  goodness of the fit of models to the observations were measured by the merit function:
\begin{equation}
MF=\frac{1}{n} \sqrt{ \overline{\sigma^2} \sum_{i=1}^{n} \frac{ \left( \nu_{o,i}-\nu_{t,i}\right)^2}{\sigma_i^2}},
\label{eq:1}
\end{equation}
where $n$ is the number of frequencies used in the fitting procedure, 
whereas $\nu_{o,i}$ and $\nu_{t,i}$ are observed and theoretical frequencies, respectively.
Finally, $\sigma_i$ stands for observed frequency errors and
$\overline{\sigma^2} = \frac{1}{n} \sum_{i=1}^{n}\sigma_i^2$.
The function $MF$, defined as above, gives approximately the average difference
between the theoretical and observed frequencies.

For a given $(\ell,\,m)$ numbers we matched theoretical oscillation spectrum to the observed one by assignment
to $\nu_1=2.95556(6)\,\mathrm{d}^{-1}$, i.e. to the mode with the highest observed amplitude,
the radial
order of theoretical
mode with the closest frequency.

The sequence from Table\,\ref{tab:S8264293} consist of 14 frequencies.
It contains modes with the largest amplitudes ($\nu_1$, $\nu_2$, $\nu_3$, $\nu_4$, $\nu_6$, $\nu_9$, $\nu_{10}$, $\nu_{11}$) that are supplemented with modes of moderate amplitudes ($\nu_{19}$, $\nu_{27}$, $\nu_{40}$, $\nu_{41}$) and of small amplitudes ($\nu_{133}$, $\nu_{156}$).
The mean period spacing is 0.0034\,d.

\subsection{The effect of various parameters}
We  fitted all frequencies from Table\,\ref{tab:S8264293}.
To this end we constructed  several extensive grids of MESA evolutionary models.
We considered all masses whose evolutionary tracks  were inside the $3\sigma$ error box of $\log T_\mathrm{eff}$
and $\log L/\mathrm L_{\sun}$ (cf. Fig.\,\ref{fig:HR_KIC8264293}).
In our calculations, the masses were sampled with a step of $0.05\,\mathrm{M}_{\sun}$
and the range was from  2.95\,$\mathrm M_{\sun}$ to  4.20\,$\mathrm M_{\sun}$.
For simplicity, we have assumed a zero-rotation in evolutionary models.
The time step and resolution of MESA models depend on numerical controls
and is not regular and so are the pulsational models. Therefore,
in order to refine and make the regular grid, the frequencies were interpolated as a function
of effective temperature with a step $\Delta \log T_\mathrm{eff}=10^{-5}$\,K. 
The main parameters of each grid are summarized in Table\,\ref{tab:grids}.

Although the rotation was ignored during evolutionary calculations, it is crucial to include the effects of rotation on high-order g mode pulsations.
It has been done  in the framework of the traditional approximation
\citep[e.g.\,][]{1970attg.book.....C,1989nos..book.....U,
1996ApJ...460..827B,1997ApJ...491..839L,2003MNRAS.340.1020T,2003MNRAS.343..125T,2007MNRAS.374..248D}
by using the linear non-adiabatic code of  \citet{2007MNRAS.374..248D}.
Here, we considered the spherical harmonic degrees, $\ell=1,\,2,\,3$ and all possible azimuthal orders, $\left|m\right|\le \ell$.
The tested range of the rotational velocity was  from 235\,$\mathrm{km\,s^{-1}}$ to 355\,$\mathrm{km\,s^{-1}}$, with the step 0.05\,$\mathrm{km\,s^{-1}}$.

\setlength{\tabcolsep}{3pt}
\begin{table*}
	\centering
	\caption{Parameters space of our model grids. In parenthesis we gave the steps. In the cases
	were the  step was variable we gave v. All grids of models except GRID\#3
	were calculated with the exponential overshooting prescription parameterized  by
	$f_\mathrm{ov}$, whereas GRID\#3 was calculated with the step
	overshooting prescription parameterized  by
	$\alpha_\mathrm{ov}$.
	}
	\label{tab:grids}
	\begin{tabular}{ccccccccc} 
\hline
GRID & $V_\mathrm{rot}$         & $X_0$ & $Z$ & $f_\mathrm{ov}/\alpha_\mathrm{ov}$ &   $\mathrm{min}\, D_\mathrm{mix}$   & opacity data\\
     &  $[\mathrm{km\,s^{-1}}]$ &       &     &                                    &  [$\mathrm{cm^2\,s^{-1}}$]          & \\
\hline
 \#1 &  $235-355(0.05)$ & 0.710 & 0.0140 & 0.0200 &   0 & OPLIB\\
 \#2 &  $245-275(0.05)$ & 0.710 & 0.0140 & 0.00-0.04(0.001) &   0 & OPLIB\\
 \#3 &  $245-275(0.05)$ & 0.710 & 0.0140 & 0.00-0.40(0.01) &   0 & OPLIB\\
 \#4 &  $245-275(0.05)$ & 0.710 & 0.0140 & 0.00-0.04(0.001) &   0 & OP\\
 \#5 &  $245-275(0.05)$ & 0.710 & 0.0140 & 0.00-0.04(0.001) &   0 & OPAL\\
 \#6 &  $245-275(0.05)$ & 0.710 & 0.0140 & 0.028 &   $0-10^4$(v) & OPLIB\\
 \#7 &  $235-275(0.05)$ & 0.710 & $0.010-0.020$(v) & 0.028 &   0 & OPLIB\\
 \#8 &  $235-275(0.05)$ & $0.68-0.74$(0.01) & 0.0115 & 0.028 &   0 & OPLIB\\
 \#9 &  $245-255(0.01)$ & $0.72$ & 0.0115 & $0.026-0.029$(0.0005) &   0 & OPLIB\\
\#10 &  $235-255(0.01)$ & $0.72$ & $0.0104-0.0118$(0.0001) & $0.0275$ &   0 & OPLIB\\
\#11 &  $235-255(0.01)$ & $0.71-0.73$(0.005) & $0.0109$ & $0.0275$ &   0 & OPLIB\\
\#12 &  $245-255(0.01)$ & $0.72-0.735$(0.005) & $0.0106-0.0113$(0.0001) & $0.0255-0.0280(0.0005)$ &  0 & O\#1\\
\hline
\end{tabular}
\end{table*}

In our first grid of models (GRID\#1) we assumed the initial chemical composition typical  for B-type stars,
i.e. $X_0=0.71$, $Z=0.014$ \citep{2012A&A...539A.143N}, the AGSS09 chemical mixture \citep{2009ARA&A..47..481A}
and OPLIB opacity tables. We used the exponentially decaying overshooting from the convective
core \citep{2000A&A...360..952H} with the efficiency parameter $f_\mathrm{ov}=0.02$.

Calculations of $MF$ (Eq.\,\ref{eq:1}) for the models from GRID\#1 and for all considered pairs of mode degrees and azimuthal orders, ($\ell$\,,$m$), allowed us
for a firm and unambiguous mode identification.
It turned out, that the frequencies given in Table\,\ref{tab:S8264293} consist of the dipole prograde modes, $\left(\ell,\,m\right)=\left(1,\,+1\right)$.
In case of these modes the lowest values of $MF$  reache  about $1 \times 10^{-3}\,\mathrm{d}^{-1}$ and for
any other combination of ($\ell$\,,$m$) we got the value of $MF$ much above $3.5 \times 10^{-3}\,\mathrm{d}^{-1}$.

\begin{figure}
	\includegraphics[angle=0, width=\columnwidth]{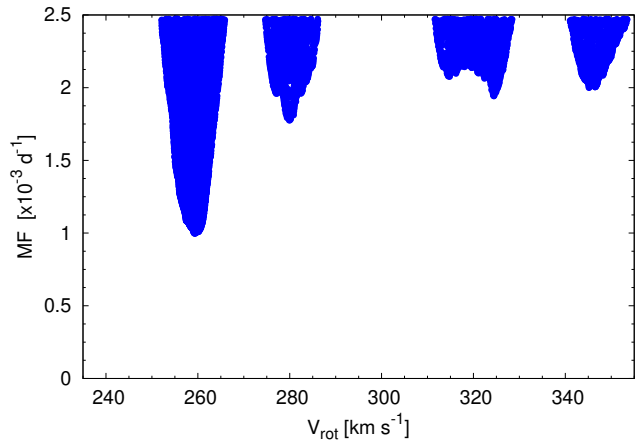}
    \caption{The values of  $MF$ as a function of the rotational velocity for  the dipole prograde modes. Models were calculated with
             $X_0=0.71$, $Z=0.014$, $f_\mathrm{ov}=0.02$ and OPLIB opacities (GRID\#1).}
    \label{fig:chi_identyfikacja}
\end{figure}

The values of $MF$ for seismic models reproducing the $(1,+1)$ modes  as a function of  the rotational velocity  are shown in
Fig.\,\ref{fig:chi_identyfikacja}. As one can see, there is a deep minimum of $MF$ for the rotational velocity
of about 260\,$\mathrm{km\,s^{-1}}$. Therefore, in the next calculations, we limited  the range of $V_{\rm rot}$ to
(245, 275)\,$\mathrm{km\,s^{-1}}$, i.e. around a vicinity  of the minimum value of $MF$.

\begin{figure}
	\includegraphics[angle=0, width=\columnwidth]{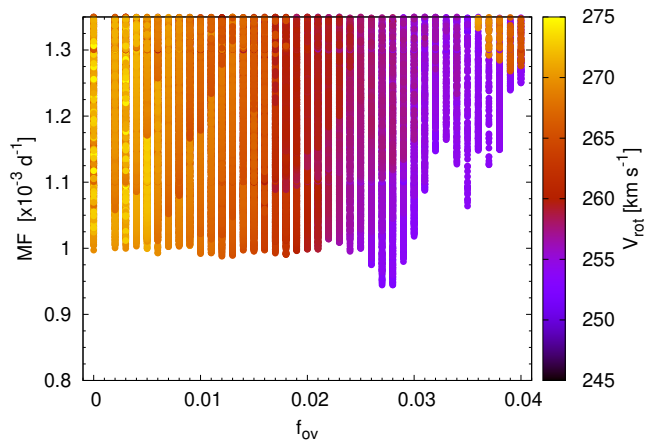}
    \caption{The values of  $MF$ as a function of the overshooting parameter, $f_\mathrm{ov}$, for  the dipole prograde modes. Models were calculated with
             $X_0=0.71$, $Z=0.014$, and OPLIB opacities (GRID\#2).  The value of the rotational velocity is colour coded. }
    \label{fig:chi_f_ov}
\end{figure}

In order to find the efficiency of the mixing processes close to the boundary of the convective core we extended our grid of models. We constructed models with
the overshooting parameter, $f_\mathrm{ov}$, in the range from 0 to 0.04 with a step 0.001 (GRID\#2).
The one exception was $f_\mathrm{ov}=0.001$. In this case MESA models did not converge.
The values of  $MF$ as a function of  $f_\mathrm{ov}$ parameter for seismic models reproducing the modes $(\ell,~m)=(1,+1)$
 are shown in Fig.\,\ref{fig:chi_f_ov}. The value of the rotational velocity is colour coded.


We found quite a flat dependence of $MF$ on $f_\mathrm{ov}$
with a weak minimum around $f_\mathrm{ov}=0.027$ and 0.028
($MF=0.944 \times 10^{-3}\,\mathrm{d}^{-1}$).
However, only for of $f_\mathrm{ov}>0.03$ the fitting is clearly  worse.
 Therefore, we set $f_{\mathrm {ov}}=0.03$ as an upper limit for the overshooting parameter.

There is also seen a clear correlation between efficiency of overshooting and rotation
velocity. In general, the higher $f_\mathrm{ov}$ parameter, the lower rotation velocity is preferred. Moreover,  the values of $MF$ increase significantly
for the lowest and the highest considered here rotation velocities.
For $V_{\rm {rot}}\lesssim 250$ kms$^{-1}$ and $V_{\rm {rot}}\gtrsim 272$ kms$^{-1}$ the merit function $MF$
is out of the scale in the figure.
This fact justifies a narrowing of the considered range of the rotation rate.

\begin{figure}
	\includegraphics[angle=0, width=\columnwidth]{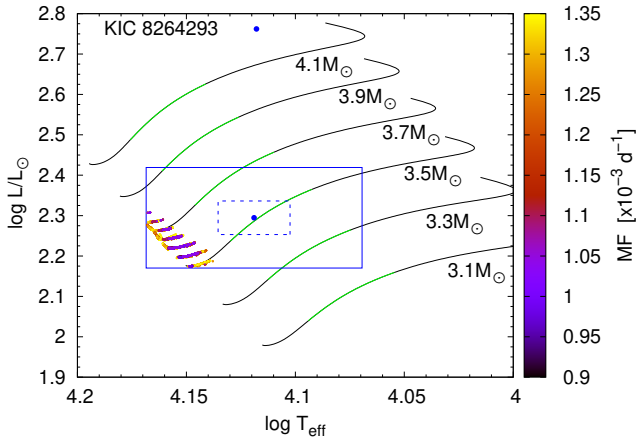}
    \caption{The same as in Fig.\,\ref{fig:HR_KIC8264293} but with the positions of seismic models from GRID\#2 with the merit function
    $MF<1.35\,\mathrm{d}^{-1}$. The values of $MF$ are colour coded. Models were calculated with
    the OPLIB opacity tables, $X_0=0.71$, $Z=0.014$ and $f_\mathrm{ov}$ from 0 to 0.04.  }
    \label{fig:HR_D_models_OPLIB_ov}
\end{figure}

The positions of the seismic models from the grid GRID\#2 with $MF<1.35\times 10^{-3}\,\mathrm{d}^{-1}$ are shown in the H-R diagram in Fig.\,\ref{fig:HR_D_models_OPLIB_ov}.
As one can see all models lie in the left bottom corner
of the $3\sigma$ error box,  close to the zero-age main-sequence (ZAMS). The central abundances
of hydrogen, $X_\mathrm{c}$, of these models are above 0.65 and reaches the value of $\sim 0.703$
for the best models. Parameters of the two best GRID\#2 models  (MODEL\#1 and MODEL\#2) are given in Table\,\ref{tab:best_models}.

\setlength{\tabcolsep}{5pt}
\begin{table*}
	\centering
	\caption{Parameters of the best seismic models in the following order: model number, mass, rotation velocity, initial hydrogen abundance, initial metallicity,
	overshooting parameter, effective temperature, luminosity, gravity, central hydrogen content, mass of the convective core,
	number of unstable modes corresponding to the fitted frequencies, minimal diffusive mixing coefficient in the radiative envelope, opacity data  and the value of the merit function.
	All models except MODEL\#3 	were calculated with the exponential overshooting prescription parameterized  by
	$f_\mathrm{ov}$. MODEL\#3 was calculated with the step 	overshooting prescription parameterized  by $\alpha_\mathrm{ov}$.}
	\label{tab:best_models}
	\begin{tabular}{ccccccccccccccc} 
\hline
 & $M$  & $V_\mathrm{rot}$  & $X_0$ & $Z$ & $f_\mathrm{ov}/\alpha_\mathrm{ov}$ & $\log T_\mathrm{eff}$ & $\log L/\mathrm{L}_{\sun}$ & $\log g$ & $X_\mathrm{c}$ & $M_\mathrm{cc}$  &  $n_\mathrm{u}$ & $\mathrm{min}\, D_\mathrm{mix}$   & opacity & $MF$ \\
 & $[\mathrm{M}_{\sun}]$ & $[\mathrm{km\,s^{-1}}]$ & & & & & & & & $[\mathrm{M}_{\sun}]$  &  & [$\mathrm{cm^2\,s^{-1}}$] & & $[\times 10^{-3}\,\mathrm{d}^{-1}]$\\
\hline
\#1 & 3.50 & 252.80 & 0.710 & 0.0140 & 0.0270 & 4.14784 & 2.1748 & 4.352 & 0.7031 & 0.77 &  3 & 0 & OPLIB &  0.944\\
\#2 & 3.50 & 252.40 & 0.710 & 0.0140 & 0.0280 & 4.14808 & 2.1747 & 4.353 & 0.7037 & 0.77 & 3 & 0 & OPLIB &  0.944\\
\#3 & 3.55 & 265.35 & 0.710 & 0.0140 & 0.20 & 4.14638 & 2.2049 & 4.322 & 0.6830  & 0.77  & 3 & 0 & OPLIB &  0.980\\
\#4 & 3.55 & 254.45 & 0.710 & 0.0140 & 0.0210 & 4.15144 & 2.1899 & 4.357 & 0.6996 & 0.79 &  5& 0 & OP & 1.044\\
\#5 & 3.50 & 251.15 & 0.710 & 0.0140 & 0.0290 & 4.15087 & 2.1811 & 4.358 & 0.7048 & 0.78 & 2 & 0 &OPAL &  1.067 \\
\#6 & 3.50 & 252.40 & 0.710 & 0.0140 & 0.0280 & 4.14808 & 2.1747 & 4.353 & 0.7036 & 0.77   & 3       & 50 & OPLIB & 0.944\\
\#7 & 3.55 & 247.50 & 0.710 & 0.0115 & 0.0280 & 4.16587 & 2.2286 & 4.377 & 0.7037 & 0.80 & 0 & 0 & OPLIB & 0.926 \\
\#8 & 3.55 & 247.80 & 0.720 & 0.0115 & 0.0280 & 4.16086 & 2.2105 & 4.375 & 0.7123 & 0.79 &  0 &  0 & OPLIB &    0.920\\
\#9 & 3.55 & 247.60 & 0.720 & 0.0115 & 0.0275 & 4.16098 & 2.2104 & 4.375 & 0.7126 & 0.80  &  0    & 0 & OPLIB  & 0.908\\
\#10 &3.55 & 246.39 & 0.720 & 0.0109 & 0.0275 & 4.16417 & 2.2190 & 4.379 & 0.7113 & 0.79 &   0& 0 & OPLIB & 0.905\\
\#11 & 3.57 & 246.61 & 0.725 & 0.0109 & 0.0275 & 4.16382 & 2.2184 & 4.381 & 0.7170 & 0.81   & 0 &    0 & OPLIB & 0.904 \\
\#12 & 3.54 & 247.88 & 0.725 & 0.0112 &  0.0260 &  4.15823 & 2.2017 & 4.372 & 0.7148&  0.80  & 14 & 0 & O\#1 & 0.909 \\
\hline
\end{tabular}
\end{table*}

\begin{figure}
	\includegraphics[angle=0, width=\columnwidth]{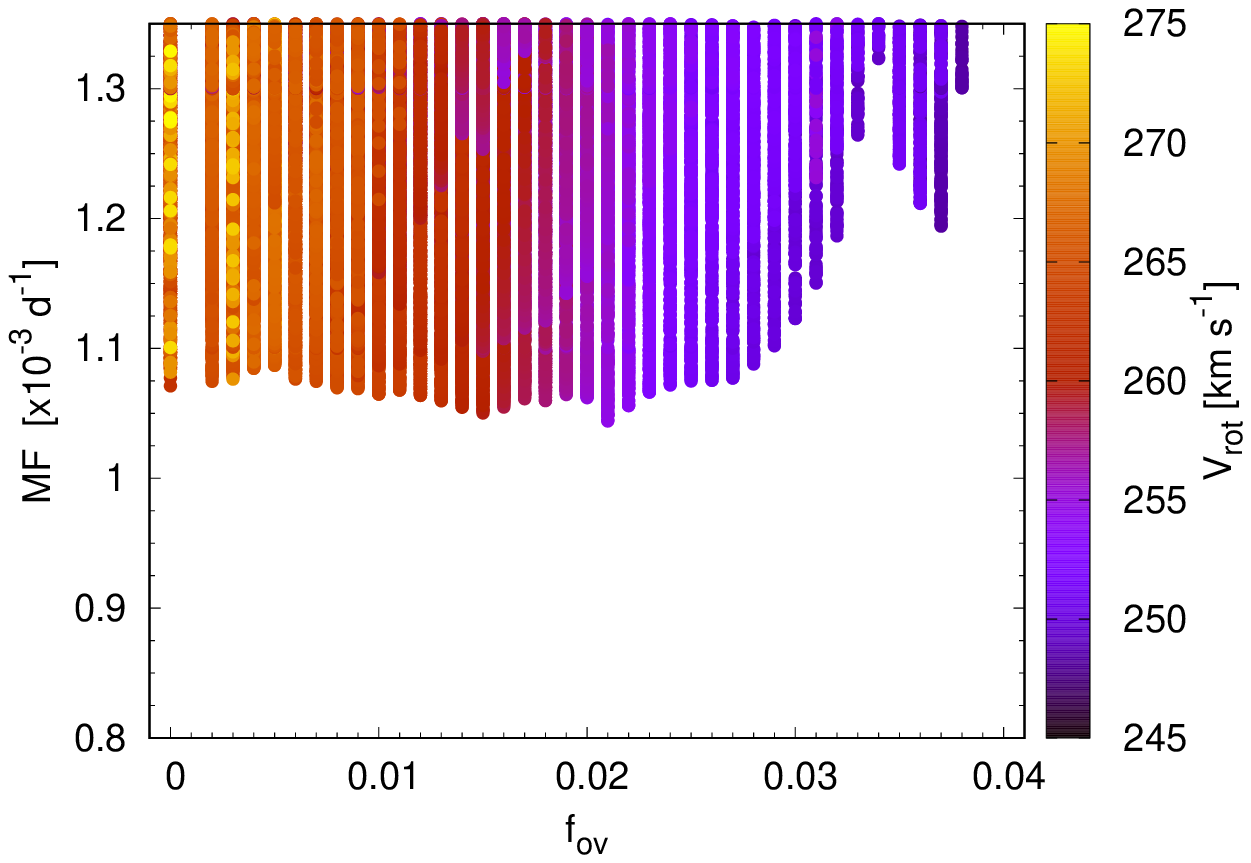}
    \caption{The same as in Fig.\,\ref{fig:chi_f_ov} but for the OP models (GRID\#4).}
    \label{fig:chi_f_ov_OP}
\end{figure}

\begin{figure}
	\includegraphics[angle=0, width=\columnwidth]{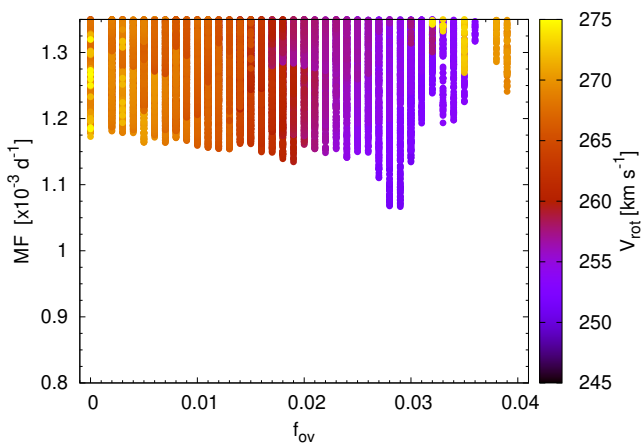}
    \caption{The same as in Fig.\,\ref{fig:chi_f_ov} but for the OPAL models (GRID\#5).}
    \label{fig:chi_f_ov_OPAL}
\end{figure}

In the next step we checked the sensitivity of the quality of the fit to the overshooting scheme used
in the calculations. Therefore we calculated new grid of models (GRID\#3 in Table\,\ref{tab:grids}) where we used
step overshooting prescription. The space of parameters in this grid was the same as in
GRID\#2 but the efficiency of overshooting was describe by the parameter $\alpha_\mathrm{ov}$
in the range  from 0 to 0.4. The best seismic model from this grid has $\alpha_\mathrm{ov}=0.2$ and $MF=0.980\times 10^{-3}\,\mathrm{d}^{-1}$
(MODEL\#3 in Table\ref{tab:best_models}). Thus, the effect is rather minor but since the values of $MF$ in GRID\#3 are slightly higher than
in GRID\#2, i.e. $MF=0.980\times 10^{-3}\,\mathrm{d}^{-1}$ for MODEL\#3 vs $MF=0.944\times 10^{-3}\,\mathrm{d}^{-1}$ for MODEL\#1 and MODEL\#2,
we decided to use only exponential overshooting scheme in further calculations.

To test the effect of opacity data we used
the OP tables \citep[][GRID\#4]{1996MNRAS.279...95S, 2005MNRAS.362L...1S} and the OPAL tables \citep[][GRID\#5]{Iglesias1993, Iglesias1996}.
As one can see from Figs.\,\ref{fig:chi_f_ov_OP} and \ref{fig:chi_f_ov_OPAL} the  values of $MF$
calculated for the OP and OPAL seismic models are higher than  those calculated for the OPLIB seismic models but
the overall run is similar, i.e., there is a kind of flattening for low values of $f_\mathrm{ov}$
and a decrease in the quality of the fit  for $f_\mathrm{ov}>0.03$. However, in the case of the OP seismic models
there is no weak minimum around $f_\mathrm{ov}=0.028$ obtained for the OPLIB and OPAL seismic models.
The propagation zone
of dipole prograde modes is located
at the depth between temperatures of about
logT=5.5 and 7.3. In this region, the OP opacities differ from the OPAL and OPLIB data, which is clearly
seen in Fig.\,1 and Fig.\,2 of \citet{2015A&A...580L...9W}.  This may be partly the reason for a lack of the small
local minimum of $MF$ at $f_\mathrm{ov}\approx 0.28$ in the case of the OP seismic models.
Parameters of  the best OP and OPAL seismic models are given in Table\,\ref{tab:best_models}  
marked as MODEL\#4 and MODEL\#5, respectively.

\begin{figure}
	\includegraphics[angle=0, width=\columnwidth]{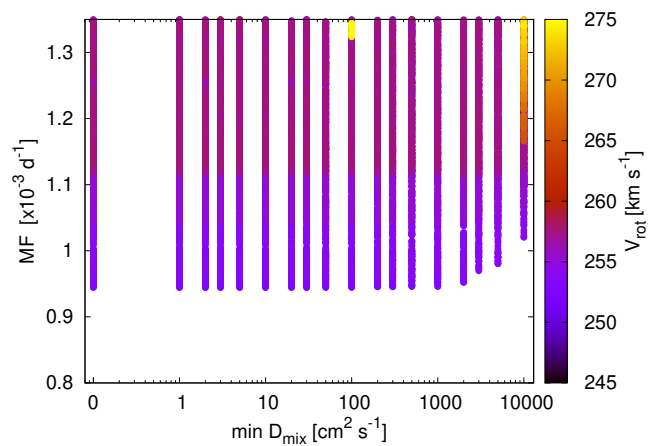}
    \caption{The values of  $MF$ as a function of the minimum mixing coefficient for the dipole prograde modes. Models were calculated with
             $X_0=0.71$, $Z=0.014$, $f_\mathrm{ov}=0.028$, and OPLIB opacities (GRID\#6). The rotational velocity is colour coded. }
    \label{fig:chi_min_Dmix}
\end{figure}

As we have mentioned at the beginning of this section we have ignored all effects of rotation in evolutionary calculations.
MESA offers the possibility of adding the rotationally induced mixing explicitly.
However, due to the fact that one needs to calibrate a few processes simultaneously, we decided
to bypass this by introducing an artificial minimal mixing. This mixing can be controlled by the minimal mixing coefficient, $\mathrm{min}\, D_\mathrm{mix}$.
In that way, we will test the efficiency of mixing in the radiative envelope that can be rotationally induced.
In MESA all mixing processes are treated in a diffusive approximation and diffusive coefficient $D$ is set to
$\mathrm{min}\, D_\mathrm{mix}$ if only $D<\mathrm{min}\, D_\mathrm{mix}$.
The similar approach was applied by \citet{2015A&A...580A..27M, 2016ApJ...823..130M}
who found a low vertical diffusive mixing in KIC\,10526294 and KIC\,7760680.

Our results so far showed that the best seismic models were obtained with the OPLIB opacities and for the overshooting parameters
$f_\mathrm{ov}=0.27$, 0.28. Therefore, in our next grid of models with a variable mixing coefficient $\mathrm{min}\, D_\mathrm{mix}$
we adopted the OPLIB tables and $f_\mathrm{ov}=0.28$ (GRID\#6).
We considered the following values of $\mathrm{min}\, D_\mathrm{mix}$: 0, 1, 2, 3, 5, 10, 20, 30, 50, 100, 200, 300, 500, 1000,
2000, 3000, 5000 and 10000 $\mathrm{cm^2\,s^{-1}}$.
The values of $MF$ as a function of $\mathrm{min}\, D_\mathrm{mix}$  are shown in Fig.\,\ref{fig:chi_min_Dmix}.

As one can see, the models with no additional mixing fit the observed frequencies from Table\,\ref{tab:S8264293}
as the dipole prograde modes
as well as the seismic  models with $\mathrm{min}\, D_\mathrm{mix}\approx 1000\,\mathrm{cm^2\,s^{-1}}$.
Only very high mixing efficiency can be excluded.
We found  the same minimum of $MF=0.944\times 10^{-3}\,\mathrm{d}^{-1}$  for the seismic models with
$\mathrm{min}\, D_\mathrm{mix}=0$, 1, 2, 3, 5, 10, 30 and 50 and 100 $\mathrm{cm^2\,s^{-1}}$.
As an example we give parameters for the model (MODEL\#6) with $\mathrm{min}\, D_\mathrm{mix}=50\,\mathrm{cm^2\,s^{-1}}$
in Table\,\ref{tab:best_models}.
Therefore,  in the subsequent calculations we decided to set  $\mathrm{min}\, D_\mathrm{mix}=0\,\mathrm{cm^2\,s^{-1}}$.
The reason of this phenomenon may be the fact that  the star is still almost on ZAMS. This means that it is actually chemically homogeneous
and additional mixing has little effect on the chemical gradients and hence pulsational properties. 

\begin{figure}
	\includegraphics[angle=0, width=\columnwidth]{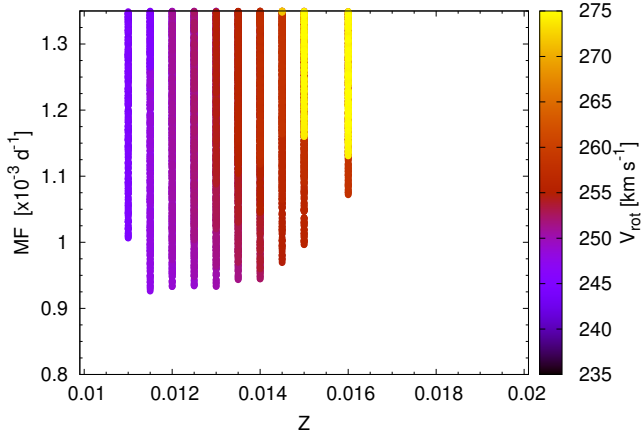}
    \caption{The values of  $MF$ as a function of $Z$ for the dipole prograde modes.
    The seismic models were calculated with $X_0=0.71$, $f_\mathrm{ov}=0.028$,
             $\mathrm{min}\, D_\mathrm{mix}=0\,\mathrm{cm^2\,s^{-1}}$
             and OPLIB opacities (GRID\#7). The rotational velocity is colour coded.}
    \label{fig:chi_Z}
\end{figure}

In the next step we studied the effect of metallicity.
We calculated a grid of models for $Z$ from 0.010 to 0.020 with the step 0.001,
 assuming $f_\mathrm{ov}=0.028$, $X_0=0.71$, $\mathrm{min}\, D_\mathrm{mix}=0\,\mathrm{cm^2\,s^{-1}}$ and the OPLIB opacity tables (GRID\#7).
We found a broad minimum in the vicinity of the solar metallicity as can be seen in Fig.\,\ref{fig:chi_Z}.
Therefore,  additional  models with $Z=0.0115,$ 0.0125, 0.0135 and 0.0145 were computed.
Since  the seismic models with a low value of the merit function were close to our lower limit for the rotational velocity
we added also calculations for $V_\mathrm{rot} \in (235,\,245)\,\mathrm{km\,s^{-1}}$. As one can see in Fig.\,\ref{fig:chi_Z},
the seismic models with the lowest value of $MF$ have metallicities approximately in the range from 0.0115 to 0.0145.
In the case of
lowest and highest considered metallicities our merit function reaches the values above $1.35\times 10^{-3}\,\mathrm{d}^{-1}$.
There is also a clear correlation between metallicity and rotation velocity: the lower  the metallicity, the lower the rotation velocity.
Parameters of the best seismic model with $Z=0.0115$ are listed in Table\,\ref{tab:best_models} (MODEL\#7).

\begin{figure}
	\includegraphics[angle=0, width=\columnwidth]{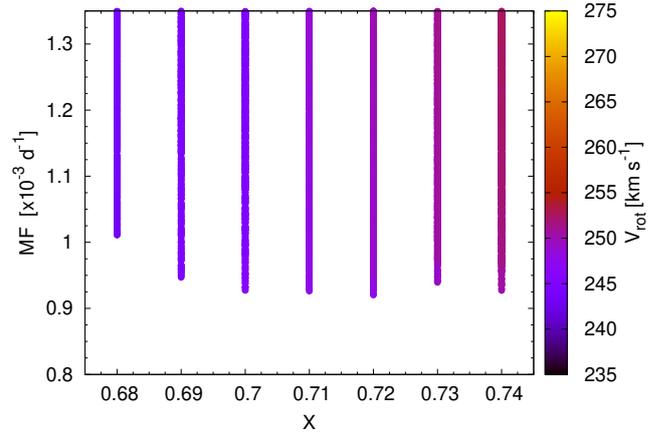}
    \caption{The values of  $MF$ as a function of $X_0$ for the dipole prograde modes . Models were calculated with
             $Z=0.0115$, $f_\mathrm{ov}=0.028$, $\mathrm{min}\, D_\mathrm{mix}=0\,\mathrm{cm^2\,s^{-1}}$
             and OPLIB opacities (GRID\#8). The rotational velocity is colour coded.}
    \label{fig:chi_X}
\end{figure}

Finally, we tested the effect of the initial hydrogen abundance, $X_0$. We considered
$X$ in the range from 0.68 to 0.74 with a step 0.01. In this grid, we used $Z=0.115$, $f_\mathrm{ov}=0.028$,
$\mathrm{min}\, D_\mathrm{mix}=0\,\mathrm{cm^2\,s^{-1}}$ and OPLIB opacity tables (GRID\#8).
For the same reasons as in the previous case we had to add calculations
for $V_\mathrm{rot}\in (235,\,245)\,\mathrm{km\,s^{-1}}$. The results are shown
in Fig.\,\ref{fig:chi_X}. The dependence of $MF$ on $X$ is weak and only the seismic models
with $X=0.68$ have clearly larger values of the merit function. However, the best
model is found for $X=0.72$ (see Table\,\ref{tab:best_models}, MODEL\#8).
There is also a clear correlation between $X$ and $V_\mathrm{rot}$: the lower the value of $X$, the lower the value of $V_\mathrm{rot}$.

\begin{figure}
	\includegraphics[angle=0, width=\columnwidth]{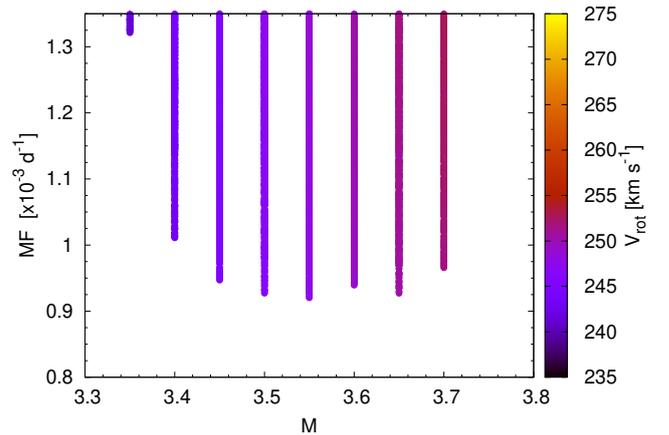}
    \caption{The values of  $MF$ as a function of the mass for the dipole prograde modes. Models were calculated with
             $Z=0.0115$, $f_\mathrm{ov}=0.028$, $\mathrm{min}\, D_\mathrm{mix}=0\,\mathrm{cm^2\,s^{-1}}$
             and OPLIB opacities  (GRID\#8). The rotational velocity is colour coded.}
    \label{fig:chi_X_masa}
\end{figure}

The value of $MF$ of the seismic models from GRID\#8 as a function of mass is shown in Fig.\,\ref{fig:chi_X_masa}.
As one can see the mass is well constrained in the range from 3.4\,M$_{\sun}$ to 3.7\,M$_{\sun}$.

Stellar rotation with the mode geometry  determine the value and slope of the period spacing.
Therefore we started our analysis from simultaneous constraining the value of rotation and mode identification.
Then we constrained $f_\mathrm{ov}$, $\mathrm{min}\, D_\mathrm{mix}$, $Z$, and $X$ one at a time.
There is some risk that such procedure
may result in finding a local minimum instead of the global minimum. However, mixing processes
near the convection core should have larger influence on the shape of period spacing than the chemical composition.
Therefore, the adopted order of searching for a solution seems to be justified. Moreover, we started from
the values of
$Z$ and $X$ typical for a star like KIC\,8326493, i.e., B-type star in the solar neighborhood \citep{2012A&A...539A.143N}.
Finally in the next section we performed fine-tuning.
The latter may allow to an exit from a possible local minimum.

\subsection{Fine-tuning of seismic models} 

\begin{figure}
	\includegraphics[angle=0, width=\columnwidth]{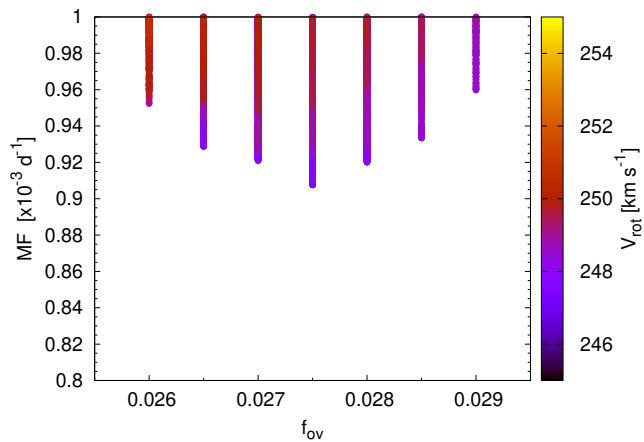}
    \caption{The values of  $MF$ as a function of $f_\mathrm{ov}$  for the dipole prograde modes. Models were calculated with
             $X_0=0.72$, $Z=0.0115$,  $\mathrm{min}\, D_\mathrm{mix}=0\,\mathrm{cm^2\,s^{-1}}$
             and OPLIB opacities  (GRID\#9). The rotational velocity is colour coded.}
    \label{fig:chi_ov_fine_tune}
\end{figure}

\begin{figure*}
	\includegraphics[angle=270, width=2\columnwidth]{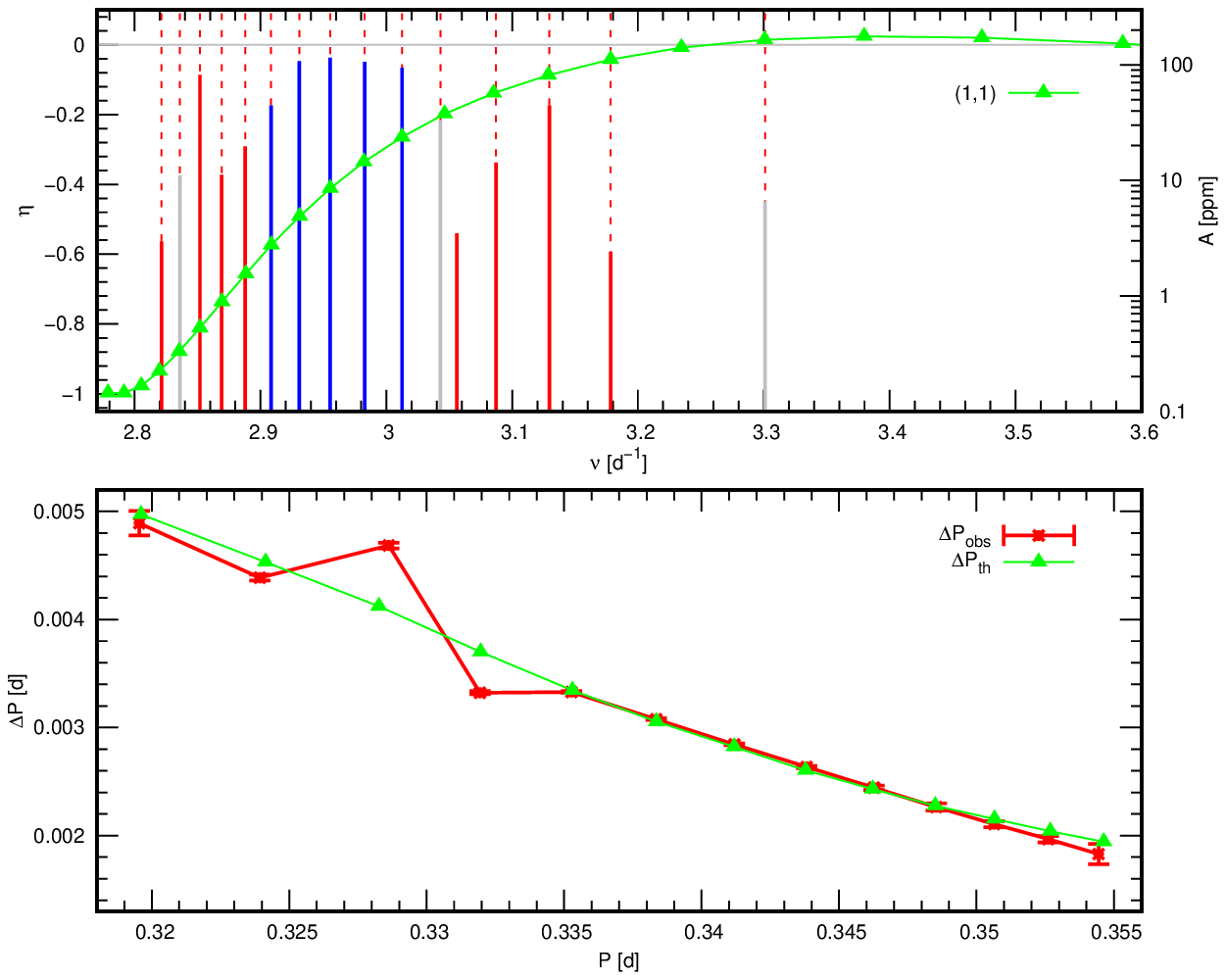}
    \caption{{\it Top panel:} The instability parameter $\eta$  (the left Y-axis) for  the  dipole prograde modes $(\ell,~m)=(1,~+1)$ as a function of
    the mode frequency for the seismic model MODEL\#11 of KIC\,8264293 . Unstable modes  are above the horizontal grey line ($\eta>0$).
    Theoretical peaks are marked with green triangles.   All observed  frequencies in the  vicinity of the g3 group are shown as peaks.
    Vertical dashed peaks indicate frequencies belonging to  the period spacing feature listed in Table\,\ref{tab:S8264293}
    and the omitted in table $\nu_{86}$.
    Independent frequencies are marked with vertical solid red peaks, whereas  combinations with vertical solid grey peaks. Vertical solid blue lines
    indicate independent frequencies for which harmonics are present. The heights of the solid peaks  correspond to the observed amplitude indicated on the right Y-axis. 
    {\it Bottom panel:} The period spacing as a function of the mode period. The observed values with the errors are marked
    with red lines and symbols, whereas the theoretical values from MODEL\#11 are marked with green lines and symbols.}
    \label{fig:eta_l1m1_stand_OPAC}
\end{figure*}

In the next step, we performed a fine tuning in the vicinity of parameters of our best seismic  model so far, i.e., MODEL\#8 (Table\,\ref{tab:best_models}).
Firstly, we decreased the steps in $M$ to 0.01\,M$_{\sun}$, in $V_\mathrm{rot}$ to 0.01\,km\,s$^{-1}$, and in $\log T_{\rm eff}$ to $10^{-6}$.
In addition, based on the previous results,
we narrowed the mass range  to (3.4,\,3.7)\,M$_{\sun}$ and the rotational velocity range to (245,\,255)\,km\,s$^{-1}$.
We proceeded in a similar way as before, i.e., we constrained the parameters one by one.

Firstly, we calculated a grid (GRID\#9) of models for $f_\mathrm{ov}=0.0260-0.0290$ with the step 0.0005,
whereas the values of $X_0$, $Z$ and $\mathrm{min}\, D_\mathrm{mix}$ were fixed as in MODEL\#8.
The main result of our parameter adjustment was a more precise value of the overshooting parameter, i.e., $f_\mathrm{ov}=0.0275$.
This new seismic model is marked as MODEL\#9 and its remaining parameters  are given in Table\,\ref{tab:best_models}.
The values of $MF$ as a function of $f_\mathrm{ov}$ for  the GRID\#9 models are shown in Fig.\,\ref{fig:chi_ov_fine_tune}.

After fixing  $f_\mathrm{ov}$ to the value of 0.0275 we tried to improve the metallicity determination. We calculated a new grid of models (GRID\#10)
spanning the metallicity in the range from 0.0104 to 0.0118 with the step 0.0001.
Once again we had to extend our calculations to the range (235,\,255)\,km\,s$^{-1}$ in the rotational velocity.
We found that the values of $MF$ reaches the minimum for $Z=0.0109$  and the corresponding seismic model is  MODEL\#10 (see Table\,\ref{tab:best_models}).

Finally, a fine tuning for the initial hydrogen abundance $X_0$ was done (GRID\#11). The values of all parameters except $X_0$ were set to their values of  MODEL\#10.
The value of  $X_0$  was search in the range from 0.71 to 0.73 with a step 0.005.
The best seismic model was found for $X_0=0.725$ and this is MODEL\#11 in Table\,\ref{tab:best_models}.

The theoretical frequencies of the  $(1,\,+1)$ modes in our best seismic model ( MODEL\#11) are compared to the observed ones
from the g3 group  in the top panel of Fig.\,\ref{fig:eta_l1m1_stand_OPAC}. As one can see all frequencies but
$\nu_{122}=3.05609(8)\,\mathrm{d}^{-1}$ in this  range belong to period spacing pattern 
found by \citet[][see also Table\,\ref{tab:S8264293}]{2021MNRAS.503.5894S}
and are
well reproduce by our model. The frequency $\nu_{122}$ has to correspond to the  mode of different degree and/or azimuthal order.
We note that $\nu_{86}=3.30087(4)\,\mathrm{d}^{-1}$ which was excluded from our fitting procedure
is also predicted by our model. However, there is a missing mode between this frequency and the last one in the sequence.

In  the top panel of Fig.\,\ref{fig:eta_l1m1_stand_OPAC} we gave also the values of
the normalized instability parameter, $\eta$ \citep{1978AJ.....83.1184S} for the dipole prograde mode (the left Y-axis) in MODEL\#11.
If the value of $\eta$  is greater than zero then the mode is unstable, that is  excited, otherwise is stable (damped) in a given model.
As one can see all  fitted modes corresponding to the observed frequencies are stable.
The only unstable  mode in this frequency range corresponds  to  $\nu_{86}$ not used to the fit.
The excitation problem concerns  more or less all our seismic models so far listed in Table\,\ref{tab:best_models}.
In the column 12 of this table we gave the number of unstable dipole prograde modes
corresponding to the observed frequencies belonging to the fitted period spacing in a given model.
As one can see, a few unstable modes in a given model were found only in MODELS\#1-\#6.
This is because of the metallicity of these models and different opacity data in the case
of MODEL\#4 and MODEL\#5. The OP models  give usually higher instability in the gravity mode region
because the Z-bump is located deeper in these opacity data.
But even in these cases most of the modes out of 14 fitted are stable.
We will try to solve this  problem in the next section.

Finally, in the bottom panel of Fig. \,\ref{fig:eta_l1m1_stand_OPAC} we compared the observed
period spacing pattern with the corresponding theoretical values for the dipole prograde modes in MODEL\#11.

\section{Opacity modified models}
\label{Sect:mod_opac}

The problem with the mode excitation occurs quite often, especially in the era of space observations that provide us with an unprecedented huge number
of frequency peaks. Usually, there is a  problem with high-order g modes
which are not excited at all, as in the case of $\beta$ Cep or $\delta$  Sct stars
\citep[e.g.,][]{2015MNRAS.452.3073B, 2019MNRAS.485.3544W},
or their instability is shifted to higher frequencies than observed
\citep[e.g.][]{2017MNRAS.466.2284D, 2017sbcs.conf..173W}.

The driving of pulsations in  B-type stars depends crucially on properties on the mean opacity profile, especially in the vicinity the Z-bump.
Therefore, to solve the problem with the g mode excitation, the opacity modification was proposed
\citep{2004MNRAS.350.1022P,2012MNRAS.422.3460S,2017MNRAS.466.2284D}
In particular, it has been shown by
\citet{2012MNRAS.422.3460S,2017MNRAS.466.2284D}
that increasing opacity at the depth corresponding to the temperature $\log T=5.46$
excites very efficiently gravity modes in a wide range of frequencies. At this temperature nickel has its maximum contribution to the Z-bump.

\begin{figure*}
	\includegraphics[angle=270, width=2\columnwidth]{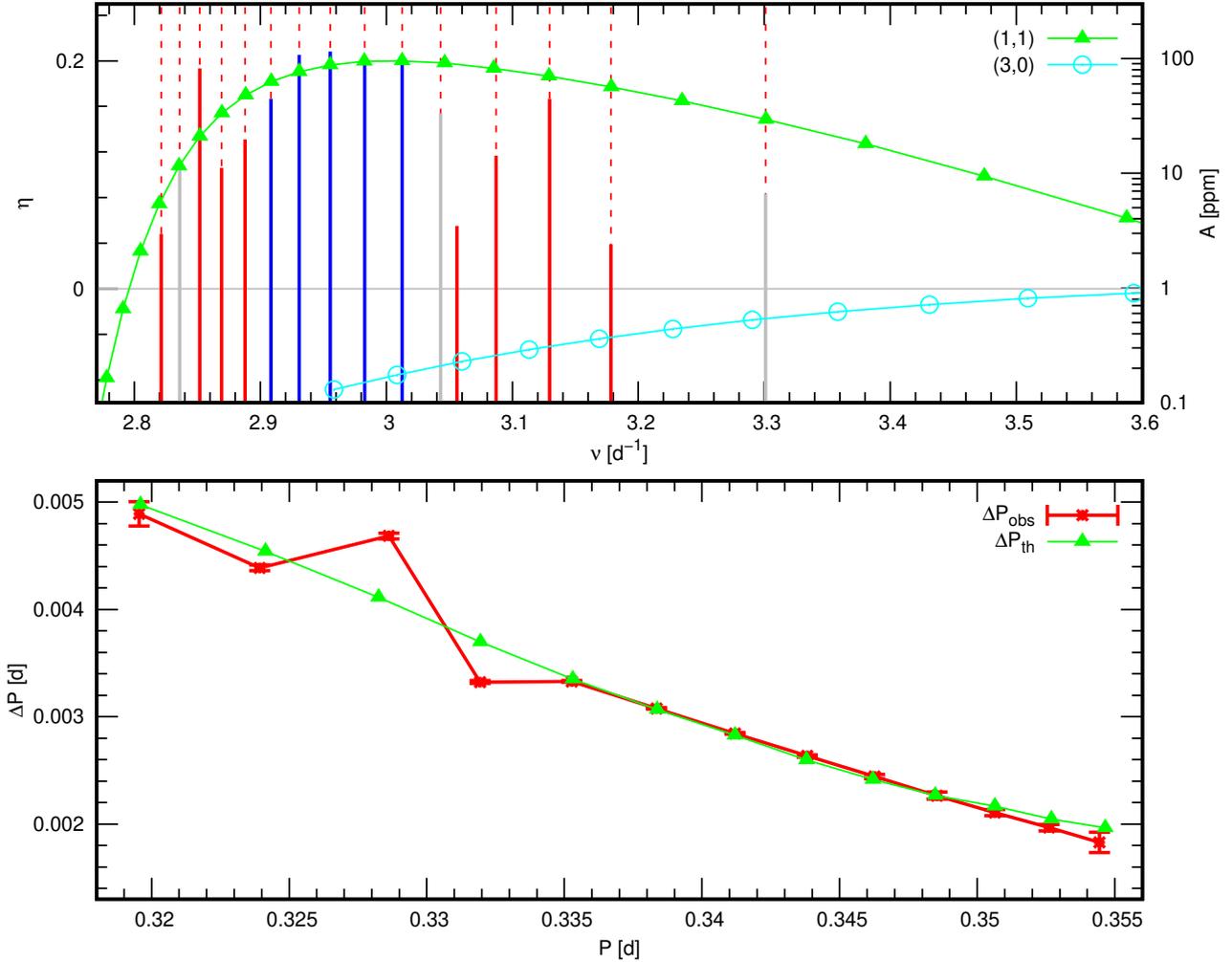}
    \caption{{\it Top panel:} The instability parameter $\eta$  (the left Y-axis) for  the (1,\,1) and (3,\,0) modes 
    as a function of
    the mode frequency for the seismic model MODEL\#12 of KIC\,8264293. Unstable modes  are above 
    the horizontal grey line ($\eta>0$).
    All observed  frequencies in the  vicinity of the g3 group are shown as peaks.
    Vertical dashed peaks indicate frequencies belonging to  the period spacing feature listed in Table\,\ref{tab:S8264293}
    and the omitted in table $\nu_{86}$.
    Independent frequencies are marked with vertical solid red peaks, whereas  combinations with vertical solid grey peaks. Vertical solid blue lines
    indicate independent frequencies for which harmonics are present. The heights of the solid peaks  correspond to the observed amplitude indicated on the right Y-axis. 
    {\it Bottom panel:} The period spacing as a function of the mode period. The observed values with the errors are marked
with red lines and symbols, whereas the theoretical values from MODEL\#12 are marked with green lines and symbols. }
    \label{fig:eta_l1_m1_stand_OPLIB_mod_opac_g3}
\end{figure*}

\begin{figure*}
	\includegraphics[angle=270, width=2\columnwidth]{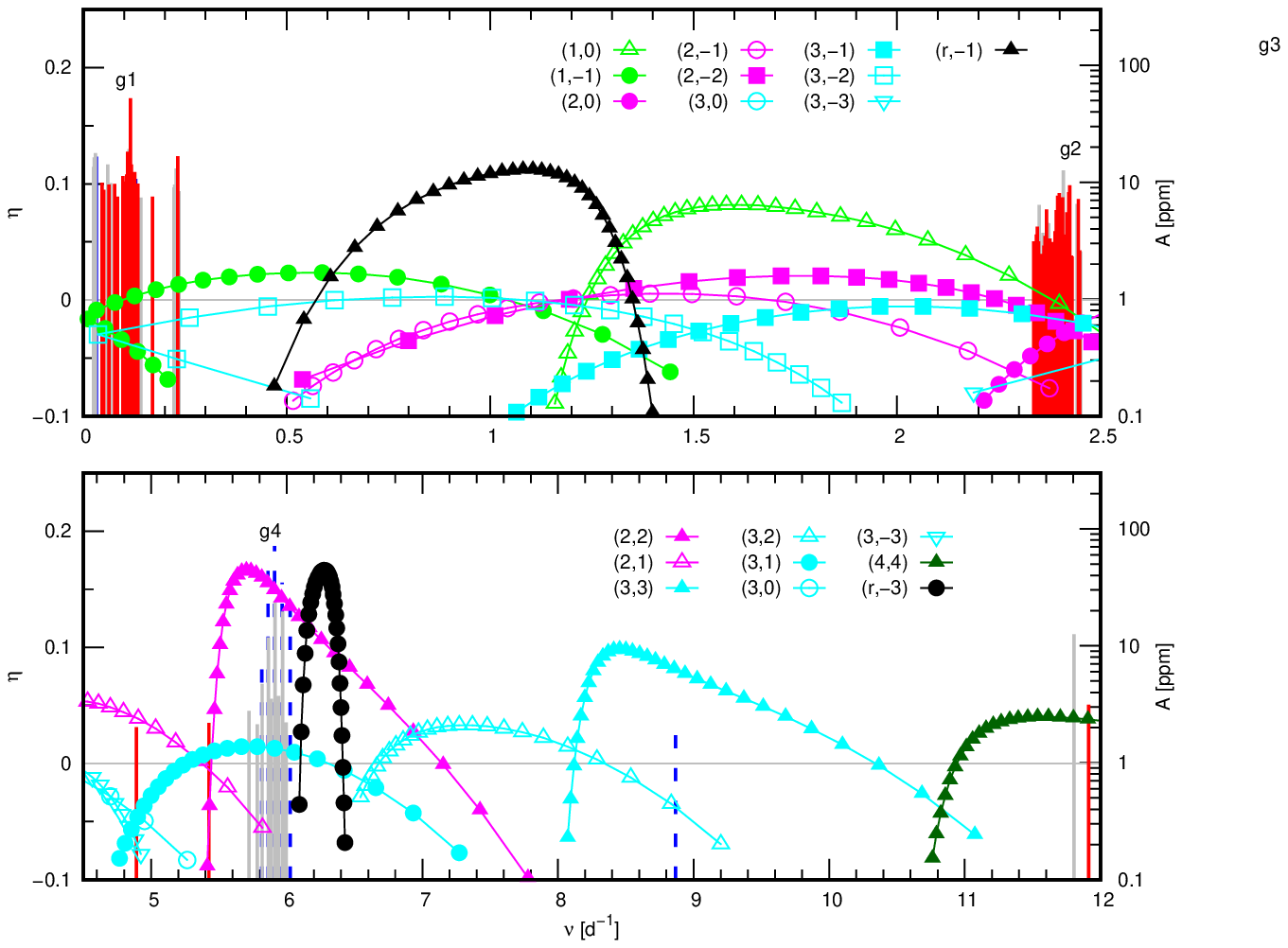}
    \caption{The instability parameter, $\eta$,   (left Y-axis) for all considered modes with $\eta >-0.1$  as a function of
    the mode frequency for seismic model of KIC\,8264293 (MODEL\#12). 
    Unstable modes  are above 
    the horizontal grey line ($\eta>0$).
    Independent frequencies are marked with vertical solid red peaks, whereas  combinations with vertical solid grey peaks.
    Vertical solid blue lines
    indicate independent frequencies for which harmonics are present and  vertical
    dashed blue lines indicate harmonics.
    The heights of the peaks  correspond to the observed amplitude indicated on the right Y-axis.
    There are shown frequency range from 0\,d$^{-1}$ to 2.5\,d$^{-1}$ (top panel)
    and 
    from 4.5\,d$^{-1}$ to 12\,d$^{-1}$ (bottom panel).}
    \label{fig:eta_l1_m1_mod259_OPLIB_2panels}
\end{figure*}

\begin{figure*}
	\includegraphics[angle=270, width=2\columnwidth]{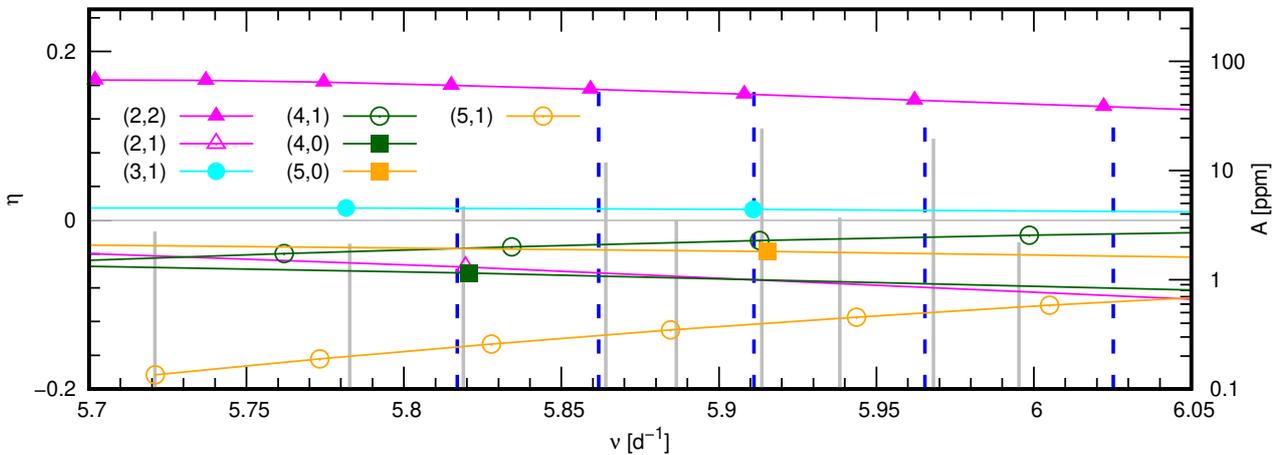}
    \caption{The instability parameter, $\eta$,   (left Y-axis) for all considered modes with $\eta >-0.2$  as a function of
    the mode frequency for seismic model of KIC\,8264293 (MODEL\#12). 
    Unstable modes  are above 
    the horizontal grey line ($\eta>0$).
    All observed  frequencies in the  vicinity of the g4 group are shown as peaks.
    Vertical solid grey peaks denote combinations, whereas
    dashed blue lines indicate harmonics.
    The heights of the peaks  correspond to the observed amplitude indicated on the right Y-axis. }
    \label{fig:eta_l1_m1_stand_OPLIB_mod_opac_g4}
\end{figure*}

\begin{figure*}
	\includegraphics[angle=270, width=2\columnwidth]{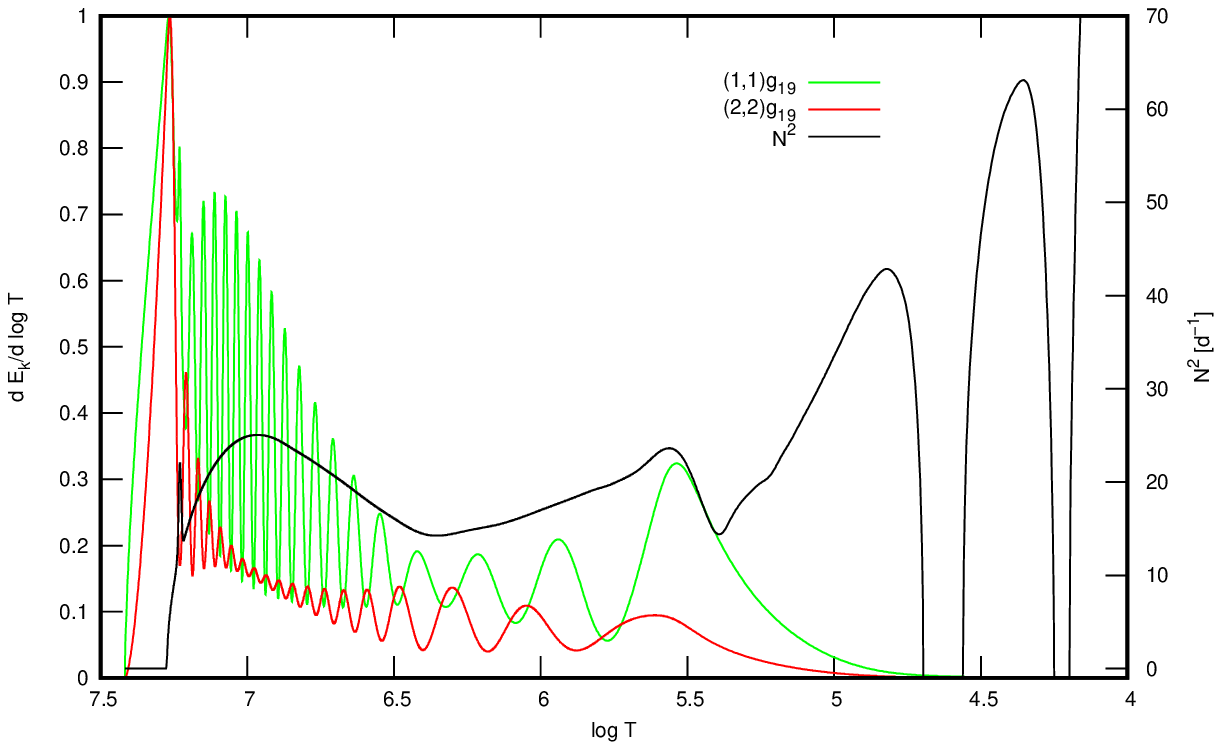}
    \caption{The differential kinetic energy for $(1,\,1)g_{19}$ and $(2,\,2)g_{19}$ modes   (left Y-axis)
    and Brunt-V\"ais\"al\"a frequency  (right Y-axis) as a function of temperature in MODEL\#12.}
    \label{fig:inertia}
\end{figure*}

Following this approach, we searched for appropriate modifications of the  opacity profile
to get instability in the whole range of frequencies belonging to the observed period spacing.
To this aim we used the method of  \citet{2017MNRAS.466.2284D} that involves
increasing or decreasing the mean opacities at certain temperatures by adding a sum of Gaussian functions in the form
\begin{equation}
\kappa\left(T\right)=\kappa_0\left(T\right)
\left[ 1+\sum_{i=1}^N b_i \exp \left( -\frac{\left(\log T - \log T_{0,i} \right)^2}{a_i^2} \right)  \right],
\end{equation}
where $\kappa_0\left(T\right)$ is the standard opacity profile.
The parameter $T_0$ stands for the central temperature at which the opacity is changed, whereas
the parameters $a$ and $b$ describe the width and height of the Gaussian profile, respectively.
$N$ gives the number of added opacity bumps.

Here, one Gaussian profile was used, i.e. $N=1$, and we considered modifications at $\log T$ in the range from 5.0 to 5.5 with a step 0.04
for $\log T<5.4$ and 0.01 for $\log T \ge 5.4$. In addition
we checked $a=0.01$, 0.02, 0.05, 0.1, 0.2, 0.3, 0.5, 0.7 and $b$ from 0.5 to 2.0 with a step 0.5.
It appeared that a number of different modifications gave appropriate instability.
We found unstable dipole prograde modes corresponding to the observed period spacing
for various modifications at $\log T$ from 5.42 to 5.5 with $a$ from 0.02 to 0.2 and $b$ from 1.0 to 2.0.

For further analysis we adopted the opacity  modification  at $\log T=5.46$ with $a=0.1$ and $b=1.0$.
This choice was driven by the fact that at this temperature nickel has its   maximum contribution to the opacity and relatively small increase of
opacity is required to obtain the proper instability. The OPLIB opacities with this modification we called O\#1.
Then  new grid of models (GRID\#12, Table\,\ref{tab:grids}) with O\#1 opacities were calculated.
We considered $M\in (3.53,~3.59)\,\mathrm{M}_{\sun}$, $V_\mathrm{rot}\in (245, ~255)\,\mathrm{km\,s^{-1}}$,
$X_0\in (0.720,~0.735)$, $Z\in(0.0106,~0.0113)$, $f_\mathrm{ov}\in (0.0255,~0.0280)$.
This parameter space cover the minimum of $MF$ found for our hitherto best MODEL\#11.

In the GRID\#12 we found the best model with $M=3.54\,\mathrm{M}_{\sun}$,
$V_\mathrm{rot}=247.88\,\mathrm{km\,s^{-1}}$,   $X_0=0.725$, $Z=0.0112$,  $f_\mathrm{ov}=0.0260$,
$\log T_\mathrm{eff}=4.15823$,    $\log L/\mathrm{L}_{\sun}=2.2017$,
and $X_\mathrm{c}=0.7148$ (see  Table\,\ref{tab:best_models}, MODEL\#12).

Next, for MODEL\#12, in addition to the dipole prograde modes, we calculated all possible pulsational modes with $\ell \le 3$,
some  higher degree modes,  $(\ell,~m)=$(4,0), (4,1), (4,4), (5,0), (5,1), as well as  the Rossby modes with $m=-1,\,-2,\,-3$.
The values of $\eta(\nu)$ in the frequency range of the g3 group
for (1,\,1) and (3,\,0) modes are depicted in   the top panel of Fig.\,\ref{fig:eta_l1_m1_stand_OPLIB_mod_opac_g3}.
As one can see all dipole prograde modes corresponding to the frequencies forming the period spacing are unstable.
These are gravity modes with the radial orders from $g_{26}$ to $g_{13}$.
The frequency $\nu_{122}=3.05609\,\mathrm{d}^{-1}$, that does not belong to the period spacing sequence,
can be associated with the mode  $(\ell,~m)=(3,0)$, but this mode has the parameter $\eta$ slightly below 0.
On the other hand, the frequency $\nu_{86}=3.30087\,\mathrm{d}^{-1}$  could belong to the period spacing sequence,
as we have already mentioned, and it would be the $g_{11}$ mode.
In the bottom panel of Fig.\,\ref{fig:eta_l1_m1_stand_OPLIB_mod_opac_g3}
we showed the period spacing pattern in our best model (MODEL\#12).

In Fig.\,\ref{fig:eta_l1_m1_mod259_OPLIB_2panels} we depicted the remaining three groups of peaks identified in the oscillation spectrum of KIC\,8264293.
The top panel includes the groups g1 and g2, whereas the bottom one the g4 group.
As one can see MODEL\#12 predicts unstable modes in these ranges of frequencies as well. Some exception is the g2 group
for which there is only one unstable mode with $(\ell,~m)=(1,0)$. But, as we explained earlier, these signals are rather 
connected with the variability in the circumstellar disk.
As for g1 group, there are unstable dipole retrograde modes in this frequency range. However, 
the theoretical spectrum is too sparse to explain all detected signals. On the other hand, at such low frequencies, below
$0.24\,\mathrm{d}^{-1}$, some signals originating in instrumental trends could survive.

MODEL\#12 predicts also unstable Rossby modes, but none of them would correspond
to the observed frequencies. Their instability cover frequencies from $0.6077\,\mathrm{d}^{-1}$ to $1.3500\,\mathrm{d}^{-1}$,
from $3.4006\,\mathrm{d}^{-1}$ to $3.8881\,\mathrm{d}^{-1}$ and from $6.1063\,\mathrm{d}^{-1}$ to $6.4079\,\mathrm{d}^{-1}$ 
for $m=-1$, -2 and -3 modes, respectively. Modes with larger $|m|$ are not likely to be detected in photometry.

In the vicinity of g4 we found the unstable prograde modes with $(\ell,~m)=(2,+2)$ and $(3,+1)$.
To reveal the exact structure of the g4 group,
we depicted this group separately in Fig.\,\ref{fig:eta_l1_m1_stand_OPLIB_mod_opac_g4}.
According to the adopted criteria, this group consists of only harmonics and combinations. However, they form quite regular
structure. Moreover,
$\nu_{62}=5.81696(4)\,\mathrm{d}^{-1}$,
$\nu_{7}=5.861829(8)\,\mathrm{d}^{-1}$,
$\nu_{5}=5.911100(6)\,\mathrm{d}^{-1}$,
$\nu_{12}=5.96536(1)\,\mathrm{d}^{-1}$ and
$\nu_{14}=6.02521(1)\,\mathrm{d}^{-1}$ are quite close to unstable prograde gravity modes $(\ell,~m)=(2,+2)$ 
with the consecutive radial orders from $n=21$ to $n=17$. The differences between $\nu_{62}$, $\nu_{7}$, $\nu_{5}$, $\nu_{12}$, $\nu_{14}$ 
and corresponding modes are $0.00192\,\mathrm{d}^{-1}$,
$0.002586\,\mathrm{d}^{-1}$,
$0.003041\,\mathrm{d}^{-1}$,
$0.00322\,\mathrm{d}^{-1}$ and $0.00303\,\mathrm{d}^{-1}$, respectively. All these theoretical frequencies are
systematically smaller than their observed counterparts. This may suggest that
the differential rotation plays a role here.

In Fig.\,\ref{fig:inertia} we showed differential kinetic energy,
$d\,E_k/d\,\log\,T$ , for dipole prograde $g_{19}$ mode and quadrupole tesseral $g_{19}$
mode.
We arbitrarily normalized it to one.
It is obvious that both modes are mostly sensitive to the conditions near the edge of the convective core. However,
high $d\,E_k/d\,\log\,T$ of quadrupole mode is more confined to the near convective core layers than in the case of dipole mode.
Since gradient of rotation is expected to occur in the transition layers between convective interior and radiative envelope
we can conclude that quadrupole modes probe layers with average rotation rate slightly higher than dipole modes do.
Therefore frequencies of our theoretical $(2,\,+2)$ modes are smaller than their observed counterparts. This discrepancy
can be removed by computing $(2,\,+2)$ modes with rotation frequency higher by only $\sim10^{-3}\mathrm{d}^{-1}$ than
rotation frequency in MODEL\#12. 

The remaining frequencies from the g4 group
can by explained by the prograde or zonal modes with $(\ell,~m)=$(2,+1), (3,+1), (4,+1), (4,0), (5,0) or (5,+1).
Therefore, frequencies from the g4 group could be independent modes that were  identified  as harmonics and combinations by chance.

Among three highest frequency signals (the bottom panel of Fig.\,\ref{fig:eta_l1_m1_mod259_OPLIB_2panels}), i.e.,
$\nu_{155}=8.8667(1)\,\mathrm{d}^{-1}$, $\nu_{33}=11.80533(3)\,\mathrm{d}^{-1}$ and
$\nu_{129}=11.91170(9)\,\mathrm{d}^{-1}$, only $\nu_{129}$ is independent according to criterion
adopted by \citet[][]{2021MNRAS.503.5894S}. However, our model predicts unstable
(3,3) modes for $\nu_{155}$ and unstable (4,4) modes for $\nu_{33}$ and $\nu_{129}$.
Moreover, the difference between $\nu_{155}$ and $g_{19}\,(3,3)$
mode is only $0.0096\,\mathrm{d}^{-1}$, the difference between $\nu_{33}$ and $g_{19}\,(4,4)$
mode is $0.00480\,\mathrm{d}^{-1}$ and the difference between $\nu_{129}$ and $g_{19}\,(4,4)$
mode is $0.00482\,\mathrm{d}^{-1}$.
This differences are only slightly higher than $MF=0.9\times 10^{-3}\,\mathrm{d}^{-1}$, i.e. the mean difference between
fitted frequencies from period spacing and corresponding dipole modes in our MODEL\#12.
The remaining two separate and independent frequencies, i.e., $\nu_{171}= 4.8894(1)\,\mathrm{d}^{-1}$
and $\nu_{165}=5.4258(1)\,\mathrm{d}^{-1}$ have also close theoretical counterparts.
The first one, $\nu_{171}$ differs from the mode $(\ell,~m)=(2,+1),~g_{12}~(\eta=0.04)$ by
$0.0160\,\mathrm{d}^{-1}$, from $(\ell,~m)=(3,+1),~g_{27}~(\eta=-0.05)$ by
$0.0105\,\mathrm{d}^{-1}$, and from $(\ell,~m)=(3,-3),~g_{29}~(\eta=-0.07)$ by
$0.0092\,\mathrm{d}^{-1}$. Whereas the frequency $\nu_{165}$ differs
from  $(\ell,~m)=(2,+2),~g_{36}~(\eta=-0.04)$ by $0.0037\,\mathrm{d}^{-1}$.


\section{Conclusions}
\label{conclusions}

The aim of this paper was to explain the light variability of the fast rotating SPB star KIC\,8264293.
Recently, \citet{2021MNRAS.503.5894S} analysed the high-precision photometric data collected in the framework of the {\it Kepler} 
space mission.
Firstly, they identified the four, very distinct groups of  frequency peaks in the observed oscillation spectrum of  KIC\,8264293.
Within the groups g3 and g4, they identified the asymptotic period spacing.
In particular, the period spacing structure within the g3 group is the most distinctive among
such structures observed in other SPB stars.
The g3 group with one exception contains  only frequencies that belong to the period spacing. 
Usually oscillation spectrum is much more noisy, i.e., in the vicinity of frequencies involved in asymptotic structure there are many other ones. 
Therefore, KIC\,8264293 gave us a unique possibility to perform firm mode identification and asteroseismic modelling based on frequencies from the g3 group.

Our simultaneous seismic modelling and mode identification, showed unambiguously that the
period spacing found in g3 group can be associated only 
 with the dipole prograde modes. Moreover, we found that the star is very young, almost still on ZAMS and with the age of about 6 Myr.
The rotation velocity is of the order of 250 km\,s$^{-1}$ what is below the observed value of $V\sin i=284(13)\,\mathrm{km\,s^{-1}}$. 
However, $V\sin i $  may be overestimated due to line broadening caused by pulsations \citep[e.g.][]{2014A&A...569A.118A,2017A&A...597A..22S}. 
Moreover, the rotation velocity we determined correspond rather to the near convective core layers than to the surface.

One of our main goals was to find constraints on the mixing efficiency  near the convective core and in radiative envelope. 
They are described by the overshooting parameter $f_{\rm ov }$ and the minimal diffusive coefficient $\mathrm{min}\, D_\mathrm{mix}$, respectively.
We obtained an upper limit on the core overshooting  $f_{\rm ov }<0.03$ whereas the addition of the mixing in the radiative
envelope  could not be constrain.   This may be because the star is very young and has not yet developed a chemical gradient.

There is also another source of variability of  KIC\,8264293. We found a weak emission in the H$\alpha$ line
which is most probably related to the Be phenomenon. We suggested that the incoherent photometric variability found in the g2 group
is caused by the fluctuation in a disk around the star.
However, more observations are needed to decide whether this is a decretion or accretion disk.
In the fist case the disk is formed from the matter ejected from the star, i.e., the classical Be object,
whereas in the second case the disk is form from the residual matter after the protostar phase. 

Finally, we examined the energetic properties of the pulsating modes, in particular their excitation.
It appeared that, in the standard opacity models, the modes corresponding to the g3 group peaks are stable.
The way to solve this problem turned out to be, as in previous works 
\citep[e.g.,][]{2017sbcs.conf..173W, 2017EPJWC.15206005W, 2017sbcs.conf..138D,2017MNRAS.466.2284D,
2018MNRAS.478.2243S, 2019MNRAS.485.3544W}, increasing the mean opacity by 100\%
at the depth corresponding to the temperature of about $\log T=5.46$.
Such opacity modification  was sufficient not only to excite the dipole prograde modes corresponding to
the observed period spacing but  to account for the instability of the whole observed oscillation spectrum.

Moreover, our best seismic model ( MODEL\#12) predicts not only unstable modes in the entire  range of the observed frequencies
but also reproduces another period spacing that was  found in the g4 group. This sequence would correspond
to the quadruple  prograde, sectoral mode, i.e., $(\ell,~m)= (2,~+2)$. Furthermore, three highest frequency peaks,
$\nu_{155}=8.8667(1)\,\mathrm{d}^{-1}$, $\nu_{33}=11.80533(3)\,\mathrm{d}^{-1}$ and $\nu_{129}=11.91170(9)\,\mathrm{d}^{-1}$,
which from
an observational point of view could be identified as low-order p/g modes typical
for $\beta$ Cephei stars, correspond in our seismic models to high-order g modes with 
$(\ell,~m)= (3,~+3)$ in the case of $\nu_{155}$ and (4,~+4) in the case of $\nu_{33}$ and $\nu_{129}$.

An interesting finding is that our model predicts theoretical
counterparts for some observed frequencies that have been identified
as combinations and harmonics. This is the case of the g4 group.
Thus, it is possible that these frequencies were identified as combinations
or harmonics only by chance.
The probability of confusing the independent frequency with a combination
increases with the number of extracted frequency peaks.
In particular, this is the case of the satellite observations.
Therefore, a simple criterion in the case of very dense oscillation spectra
cannot be taken as the final condition for recognizing the frequency as independent or not.

KIC 8264293 is a very interesting, fast-rotating, young pulsator
and certainly deserves further observations and
studies. In particular, more spectroscopic data should be collected
in order to investigate the variability of emission features in the Halpha
line and to define their nature. 

\section*{Acknowledgements}

This work was supported financially by the Polish National Science Centre grant 2018/29/B/ST9/01940.
Calculations were carried out using resources provided by Wroclaw Centre for
Networking and Supercomputing (http://wcss.pl),  grant no.  265.
This paper includes
data collected by the Kepler mission and obtained from the MAST
data archive at the Space Telescope Science Institute (STScI).
Funding for the Kepler mission is provided by the NASA Science Mission
Directorate. STScI is operated by the Association of Universities
for Research in Astronomy, Inc., under NASA contract NAS 5-26555.
Based on observations obtained with the Apache Point Observatory
3.5-meter telescope, which is owned and operated
by the Astrophysical Research Consortium. We thank Jason Jackiewicz for his support during observations on ARCES.

\section*{Data availability}

All the data underlying this paper
will be shared on reasonable request to the corresponding author.
We make an inlist needed to reproduce our
MESA results publicly available at Zenodo. These can be
downloaded at
https://doi.org/10.5281/zenodo.5872424.




\bibliographystyle{mnras}
\bibliography{KIC8264293} 




\bsp	
\label{lastpage}
\end{document}